\documentclass{article} 
\usepackage{iclr2027_conference}
\usepackage{times}

\usepackage{amsmath,amsfonts,bm}

\def\eqref#1{equation~\ref{#1}}

\def\1{\bm{1}}

\DeclareMathAlphabet{\mathsfit}{\encodingdefault}{\sfdefault}{m}{sl}
\SetMathAlphabet{\mathsfit}{bold}{\encodingdefault}{\sfdefault}{bx}{n}

\usepackage[most]{tcolorbox}
\tcbuselibrary{listings,skins}

\usepackage[a-1b]{pdfx}

\usepackage{framed}
\usepackage{siunitx}
\usepackage{latexsym}
\usepackage{colortbl}
\usepackage{xcolor}
\usepackage{nicefrac}
\usepackage{booktabs}
\usepackage{fnpct}
\usepackage{amsfonts}
\usepackage[T1]{fontenc}
\usepackage{bold-extra}
\usepackage{amsmath}
\usepackage{amssymb}
\usepackage{bm}
\usepackage{graphicx}
\usepackage{mathtools}
\usepackage{microtype}
\usepackage{multirow}
\usepackage{multicol}
\usepackage{xpatch}
\usepackage{latexsym,comment}
\usepackage[normalem]{ulem}

\usepackage{algorithm}
\usepackage{algpseudocode}
\usepackage[inline]{enumitem}
\usepackage{overpic}
\usepackage{subcaption}
\usepackage{tabularx}
\usepackage{xspace}

\makeatletter
\xpatchcmd{\@setref}{\bfseries}{\missingreference}{}{}
\def\@citex[#1]#2{\leavevmode
    \let\@citea\@empty
    \@cite{\@for\@citeb:=#2\do
        {\@citea\def\@citea{,\penalty\@m\ }%
            \edef\@citeb{\expandafter\@firstofone\@citeb\@empty}%
            \if@filesw\immediate\write\@auxout{\string\citation{\@citeb}}\fi
            \@ifundefined{b@\@citeb}{\hbox{\reset@font\missingcitation}%
                \G@refundefinedtrue
                \@latex@warning
                {Citation `\@citeb' on page \thepage \space undefined}}%
            {\@cite@ofmt{\csname b@\@citeb\endcsname}}}}{#1}}
\makeatother

\usepackage{csquotes}

\newcommand{\graph}[0]{$\mathcal{C}$\xspace}
\newcommand{\start}[0]{$\mathcal{C}_\text{init}$\xspace}
\newcommand{\edit}[0]{$\mathcal{C}_\text{edit}$\xspace}
\newcommand{\pred}{$\mathcal{C}_{\text{pred}}$\xspace}
\newcommand{\true}{$\mathcal{C}_{\text{true}}$\xspace}

\newcommand{\model}[0]{\textsc{Cabra}\xspace}

\newcommand{\gem}[1]{\mbox{\textsc{gem}}}
\newcommand{\abr}[1]{\textsc{#1}\xspace}

\newcommand{\hidetext}[1]{}
\newcommand{\ignore}[1]{}

\newtcbox{\inlinecode}{
  on line,
  colback=black!8,
  colframe=black!8,
  boxrule=0pt,
  arc=1pt,
  boxsep=1pt,
  left=2pt,
  right=2pt,
  top=1pt,
  bottom=1pt,
  fontupper=\ttfamily
}

\newif\ifcomment
\commentfalse
\ifcomment
    \newcommand{\pinaforecomment}[3]{\colorbox{#1}{\parbox{.8\linewidth}{#2: #3}}}

    \newcommand{\prtodo}[1]{\pinaforecomment{lightblue}{pr}{#1}}
    \newcommand{\prtodoi}[1]{\pinaforecomment{lightblue}{pr}{#1}}
\else
    \newcommand{\pinaforecomment}[3]{}
    \newcommand{\prtodo}[1]{}
    \newcommand{\prtodoi}[1]{}
\fi

\newcommand{\smallurl}[1]{ \begin{tiny}\url{#1}\end{tiny}}

\definecolor{lightblue}{HTML}{3cc7ea}
\definecolor{CUgold}{HTML}{CFB87C}
\definecolor{grey}{rgb}{0.95,0.95,0.95}
\definecolor{ceil}{rgb}{0.57, 0.63, 0.81}
\definecolor{UMDred}{HTML}{ed1c24}
\definecolor{UMDyellow}{HTML}{ffc20e}

\definecolor{bashbg}{HTML}{F6F8FA}
\definecolor{bashborder}{HTML}{D0D7DE}
\definecolor{codeblockbg}{HTML}{F7F7FF}
\definecolor{codeblockframe}{HTML}{000080}
\definecolor{promptbg}{HTML}{F7F7F7}
\definecolor{promptframe}{HTML}{666666}

\newtcblisting[auto counter,number within=section]{codeblock}[1][]{
    enhanced,
    breakable,
    listing only,
    colback=codeblockbg,
    colframe=codeblockframe,
    colbacktitle=codeblockframe,
    coltitle=white,
    boxrule=0.8pt,
    arc=3pt,
    outer arc=3pt,
    left=6pt,
    right=6pt,
    top=6pt,
    bottom=6pt,
    toptitle=3pt,
    bottomtitle=3pt,
    listing options={
        language=Python,
        basicstyle=\ttfamily\small,
        breaklines=true,
        columns=fullflexible,
        keepspaces=true,
        showstringspaces=false
    },
    #1
}

\newtcblisting[auto counter,number within=section]{prompt}[1][]{
    enhanced,
    breakable,
    listing only,
    colback=promptbg,
    colframe=promptframe,
    colbacktitle=promptframe,
    coltitle=white,
    boxrule=0.8pt,
    arc=3pt,
    outer arc=3pt,
    left=6pt,
    right=6pt,
    top=6pt,
    bottom=6pt,
    toptitle=3pt,
    bottomtitle=3pt,
    listing options={
        basicstyle=\ttfamily\small,
        breaklines=true,
        columns=fullflexible,
        keepspaces=true,
        showstringspaces=false
    },
    #1
}

\usepackage{hyperref}
\usepackage{cleveref}
\usepackage{url}
\usepackage{soul}

\crefformat{section}{\S#2#1#3}
\Crefformat{section}{\S#2#1#3}
\crefformat{subsection}{\S#2#1#3}
\Crefformat{subsection}{\S#2#1#3}
\crefformat{subsubsection}{\S#2#1#3}
\Crefformat{subsubsection}{\S#2#1#3}

\definecolor{redcolor}{RGB}{236,170,136}
\definecolor{bluecolor1}{RGB}{222,234,246}
\definecolor{bluecolor2}{RGB}{160,220,246}
\definecolor{bluecolor3}{RGB}{206,238,250}

\newcommand{\hlblueone}[1]{%
  \sethlcolor{bluecolor1}\hl{#1}%
}
\newcommand{\hlbluetwo}[1]{%
  \sethlcolor{bluecolor2}\hl{#1}%
}
\newcommand{\hlbluethree}[1]{%
  \sethlcolor{bluecolor3}\hl{#1}%
}

\newcommand{\hlred}[1]{%
  \sethlcolor{redcolor}\hl{#1}%
}

\title{Code Understanding is a Bottleneck for\\Coding Agents}

\author{
Nishant Balepur$^{1, 2}$ \quad Kiran Tomlinson$^{1}$ \quad Tobias Schnabel$^{1}$ \\[0.5em]
$^{1}$Microsoft Research \qquad $^{2}$University of Maryland
}

\begin{document}

\iclrfinalcopy
\maketitle
\lhead{Preprint}

\begin{abstract}
Repository benchmarks (e.g., SWE-bench) for coding agents often assume that~lines of code edited can predict task difficulty, but such datasets' poor control over code and task types makes it hard to know which abilities truly drive agent errors.
We~present~\model{}: a Coding Ability~Blueprint for Rigorous Agent evaluation.
\model{} builds tasks from scratch as call graph transformations and scales difficulty via a task size parameter on four~axes:~function traversal, search, runtime resolution, and instruction following.
We run eight LLMs and six coding agents on~6,840 \model{} tasks to show: 
1) LLM~accuracy falls as task size~grows, but agents stay near-perfect by offloading work to tools (e.g., \inlinecode{grep});
2) Larger \model{} tasks elicit more tool calls for reading and analysis, while a separate study on SWE-bench Verified shows these tool call counts predict agents' accuracy better than lines of code edited---suggesting task difficulty for agents can lie in \textit{understanding} code to edit, not just in making edits;
3) Extending \model{} to an intense understanding task~where models analyze divergent logic across two classes backs this finding, as~agent accuracy finally falls. 
More broadly, we argue for synthetic evaluations like \model{} to unmask LLM weaknesses trivialized~by tools (e.g., needle-in-a-haystack) and abilities beyond just editing (e.g., understanding) that coding~agents still~find difficult---pairing SWE-bench-style tasks~with~controlled diagnosis.\footnote{We release our code at: \url{https://github.com/microsoft/CABRA}}
\end{abstract}

\section{Introduction: Herding Coding Agent Abilities with \model{}} \label{section:intro}

Coding agents execute user queries~by reasoning about, editing, and running code \citep{chen2021codex}.
They are often evaluated on benchmarks where researchers collect GitHub repositories,~then~augment them with problem statements and test cases as issue resolution tasks \citep{jimenez2024swebench}.

Progress in coding agents requires knowing when agents fail, but repository-augmented benchmarks impede this.
Tasks jointly differ in their code and task types, supporting only~correlational analyses \citep{mehtiyev2026beyond} that can misidentify why agents~fail.
For example, \citet{deng2026swebench} attribute agents' low accuracy in SWE-Bench Pro to agents~editing more code, but our analysis shows that agents also have low accuracy when they read, analyze, and test~their~codebases~more~(\cref{subsection:swe_bench}).

To escape such confounders, researchers have long relied on synthetic tasks that isolate model abilities, enabling evaluations of how~model~performance on these abilities scales based on task size~\citep{merrill-etal-2020-formal}.
For~example, Needle in a Haystack \citep{kamradt2023needle} offers a controlled evaluation for long-context reasoning over adjustable corpus~sizes.
While prior work has extended this methodology to reasoning LLMs \citep{motwani2026longcot} and RL environments~\citep{stojanovski2026reasoning}, it has not yet been~studied in coding agents with vastly different behavior, thanks~to~their ability to call tools.

We present \model: a \textbf{C}oding \textbf{A}bility \textbf{B}lueprint for \textbf{R}igorous \textbf{A}gent evaluation, a framework for generating programming tasks from scratch.
Unlike repository-based benchmarks that derive tasks from codebases, \model{} splits task design into two controllable steps: defining task types and defining how those tasks scale in size.
We instantiate these two steps by~generating programs from directed acyclic call graphs and transforming them into five task types~(Figure~\ref{fig:cabra} left, \cref{subsection:dag}), then scaling those tasks across four axes synthesized from prior work: function traversal, function search, runtime resolution, and instruction following (Figure~\ref{fig:cabra} right,~\cref{subsection:complexity}).

We generate 6,840 \model{} task instances of increasing sizes (\cref{subsection:task_generation}), while controlling for program and~task~type.
By varying one scaling axis at a time across many codebases and tasks, we can~robustly~surface axes that are hardest for models.
We run eight LLMs and six coding agents in Copilot's public harness (\cref{subsection:agent}) on these tasks to discover how our scaling axes impact task accuracy for LLMs without tools versus agents with them (\cref{subsection:metrics}), and reveal opportunities for improving~agents.

While LLM accuracy falls on larger \model{} tasks, agents can sustain near-perfect scores~by offloading work to tools, like \inlinecode{grep} to search long code (\cref{subsection:compare_agent_llm}).
To~dig deeper, we label agents' tool calls: to read, analyze, search, edit, or~test~code (\cref{subsection:reasoning_trace}).
As task size grows, agents do~not always edit more, contrary to prior explanations for low accuracy \citep{deng2026swebench}.~Instead, they often spend more tokens on tools to read and analyze code---\textit{understanding} code to edit.
We further verify that this behavior, discovered using \model{}'s synthetic tasks, generalizes to more realistic code editing tasks: analyses on SWE-bench Verified traces reveal that the number of tool calls an agent spends to understand code predicts task difficulty better than line edit count~(\cref{subsection:swe_bench}).

Based on our analyses, we hypothesize~that understanding-intensive tasks with relatively few edits can challenge agents.
To causally test this hypothesis, we extend \model{} to a new task that targets and controls the amount of code that agents must understand: given two~classes, extract all shared logic~and~preserve any differences (\cref{subsection:understanding_task}).
When task size~grows, agent~accuracy~falls, tokens spent on understanding tools balloon, and edits stay small---supporting code understanding as a weakness (\cref{subsection:understanding_results}).
Follow-up~analyses suggest agents fail to reason over~logical decomposition, over-rely on surface-level naming, and struggle to excise one-line differences.

Our paper has implications for both researchers and developers.
For researchers, we argue for more work in controllable tasks like \model{} beyond SWE-bench-style tasks to attribute agent errors.
For developers, we highlight code understanding as a key accuracy and efficiency bottleneck, suggesting a direction for future gains. To summarize, our core contributions are:

\begin{figure}
    \centering
    \begin{overpic}[width=\linewidth]{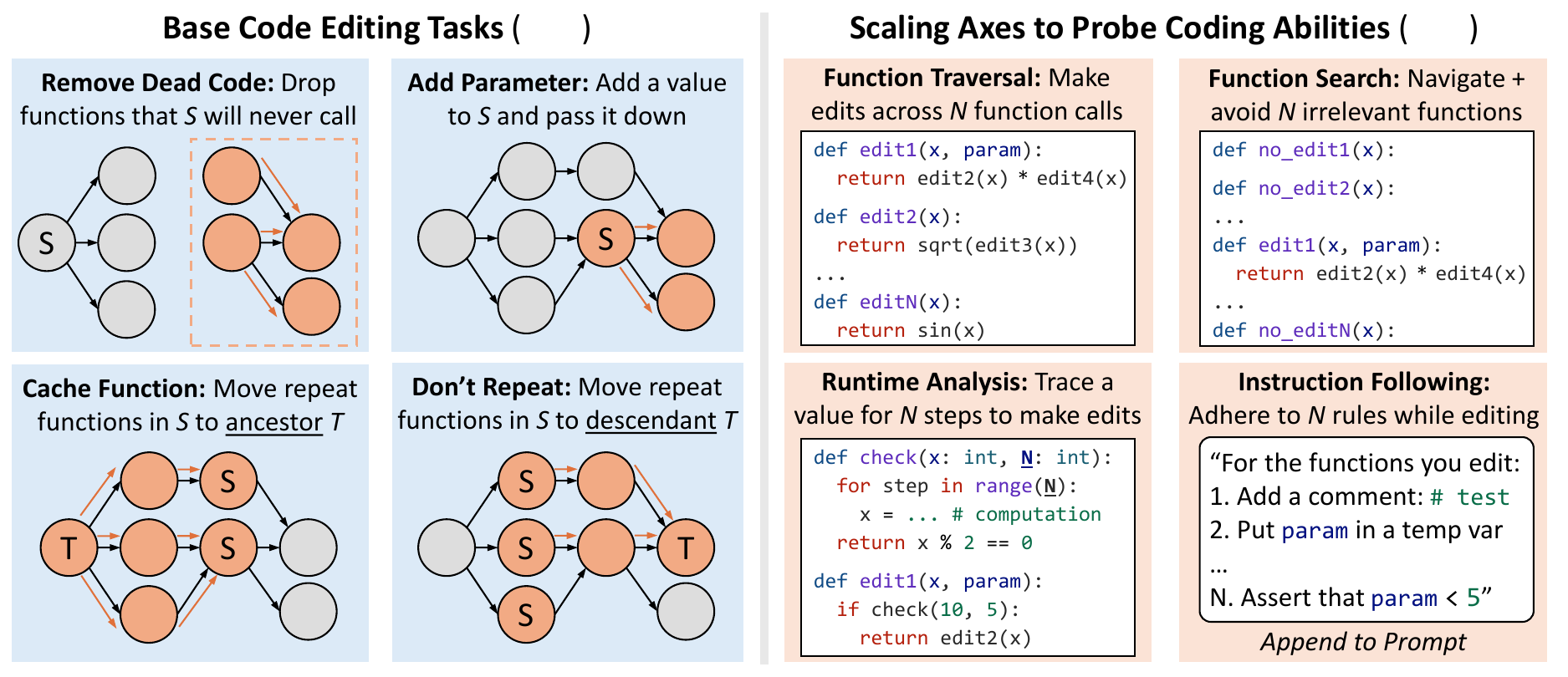}

    \put(33.2,40.7){\scalebox{0.84}{\cref{subsection:dag}}}


    \put(89.65,40.7){\scalebox{0.84}{\cref{subsection:complexity}}}

    \end{overpic}
    \vspace{-4.25ex}
    \caption{Overview of \raisebox{-0.15em}{\includegraphics[height=1.1em]{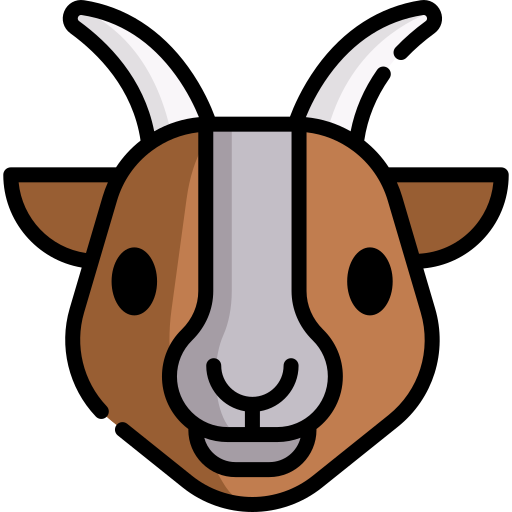}} \model{}, a framework for generating controlled coding tasks in two steps.
    We first transform directed acyclic call~graphs to define base tasks (left). We then~scale each task~over four axes via task size $N$ to probe abilities (``Add Parameter''~example, right).}
    \label{fig:cabra}
    \vspace{-1.25ex}
\end{figure}




\begin{enumerate}
[itemsep=4pt, topsep=0pt, parsep=0pt, partopsep=0pt]
    \item \model{}, a framework that generates coding tasks by sampling task~types and scaling their size, affording more controlled analyses than repository-derived benchmarks.
    \item Evidence that once-elusive tasks for LLMs are now trivial for agents that execute~code, and that code understanding is a scalable ability beyond edit volume to heighten task difficulty.
    \item A challenging code merging task to help researchers track agents' code understanding ability.
    \item A taxonomy of agent tool calls  and classifier to support future coding agent analyses.
\end{enumerate}

\section{\model{}: Taking Control of Coding Agent Evaluations} \label{section:static}

To probe coding agent failures, we present \model{}: a \textbf{C}oding \textbf{A}bility \textbf{B}lueprint for \textbf{R}igorous \textbf{A}gent evaluation.
Rather than augmenting repositories \citep{jimenez2024swebench}, \model{}~creates tasks from scratch via two controlled steps: defining tasks (\cref{subsection:dag}) and scaling task size~(\cref{subsection:complexity}). 

\subsection{Transforming Directed Acyclic Graphs as Code Editing Tasks} \label{subsection:dag}

To control our baseline code editing tasks, we use directed acyclic graphs (DAGs) to represent programs, and then transform them to define code edits.
Each DAG \graph $= \langle V, E\rangle$ is a function call graph, where $v_i \in V$ is a function and edge $(v_i, v_j) \in E$ means $v_i$ calls $v_j$. Each function computes a random mathematical function of its input.
For $v_i$ with many output vertices, $v_i$ chains the function outputs with random operations.
Appendix~\ref{appendix:sample_tasks} describes the sampling process.
Figure~\ref{fig:cabra} has example DAGs/programs with floats; we also design string and array versions so our analyses generalize beyond a single code type (Appendix~\ref{appendix:example_tasks} has examples).
We finally add a \inlinecode{\_\_main\_\_} block calling the root vertex in $\mathcal{C}$. This synthetic code generation approach trades downstream realism for simplicity and tight control over tested coding abilities.

We define five diverse tasks based on these DAGs to ensure our later task scaling analyses generalize beyond a single task type.
We ground the tasks in \citet{fowler2018refactoring}'s catalogue of refactoring types, also used to derive \citet{gautam2025refactorbench}'s tasks in RefactorBench.
Each task samples a graph \start that models do not edit and a graph \edit that models must edit, combined into the input codebase $\mathcal{C}$.
We outline our tasks next, visualized in Figure~\ref{fig:cabra} (left).

\begin{enumerate}[
    leftmargin=*,
    labelsep=0.4em,
    itemindent=0pt,
    topsep=0pt,
    leftmargin=1em,
    itemsep=0.3em,
    parsep=0pt
]
    \item \textbf{Remove Dead Code:}
    We sample \edit as a DAG representing functions that \graph never calls from \inlinecode{main}. The agent must remove all of the ``dead'' functions that exist in \edit.

    \item \textbf{Add Parameter:}
    We sample \edit as a DAG. Its one root function adds an input~parameter (e.g., a multiplier) that must be passed down the call graph and applied at~child nodes.

    \item \textbf{Add Return:}
    (2)'s inverse: we flip \edit's edges to have one sink node. The sink function adds a
    return value (e.g., a string key) that must be passed all the way up the call graph.

    \item \textbf{Cache Function:}
    We sample \edit as a tree~and call \inlinecode{slow()} at child nodes. Models move \inlinecode{slow()} to the lowest common ancestor and pass its value down to cut redundancy.

    \item \textbf{Don't Repeat:}
    The inverse of (4): we flip \edit's edges, then at each child node, we~call \inlinecode{get(val)} and apply its output. Agents~move \inlinecode{get} to the first common descendant and pass
    its value up, i.e., the \textsc{DRY} principle \citep[\textbf{D}on't \textbf{R}epeat \textbf{Y}ourself]
          {andrew2000pragmatic}.
\end{enumerate}

\noindent We sample edges between \start and \edit with probability $\beta = 0.3$ (except in Remove Dead Code, or \edit is no longer dead code).
For each task, we develop a problem statement~that explains how to solve the task with an example (all problem statements in Appendix~\ref{appendix:prompts}).





\begin{table}[t]
\centering
\caption{Four of the coding agent abilities we study, each scaled by a task size parameter~$n$.}
\label{table:complexity}
\vspace{-1.5ex}
\small
\begin{tabularx}{\linewidth}{@{}
    >{\raggedright\arraybackslash}p{0.19\linewidth}
    >{\raggedright\arraybackslash}p{0.43\linewidth}
    >{\raggedright\arraybackslash}X
@{}}
\toprule
\textbf{Scaling Param $n$} & \textbf{Ability} & \textbf{Description} \\
\midrule
\# Edit Funcs.
& Function Traversal \citep{wang2023can}
& Making $n$ edits over call graphs. \\

\# Non-edit Funcs.
& Function Search \citep{kamradt2023needle}
& Avoiding $n$ distractor functions. \\

\# Runtime Steps
& Runtime Resolution \citep{gu2024cruxeval}
& Resolving values after $n$ steps. \\

\# Instructions
& Instruction Following \citep{pyatkin2026generalizing}
& Satisfying $n$ rules while editing. \\
\bottomrule
\end{tabularx}
\vspace{-1ex}
\end{table}

\subsection{Scaling Task Size Across Complexity Axes} \label{subsection:complexity}

We now scale the size of our base tasks on four axes (Figure~\ref{fig:cabra}, right).
Each axis stress-tests a coding ability by scaling one aspect of the task via a size parameter $n$.
Our axes do not cover all abilities that agents need, but have been studied in prior work and argued as useful for downstream coding tasks.
As \cref{section:understanding} shows, \model{} can be readily extended to other coding abilities of researchers' interest.
The main scaling axes we define are (summary in~Table~\ref{table:complexity}):

\begin{enumerate}[
    leftmargin=*,
    labelsep=0.3em,
    itemindent=0pt,
    topsep=0pt,
    leftmargin=1em,
    itemsep=0.5em,
    parsep=0pt
]
    \item \textbf{Function Traversal:} Tasks such as refactoring often require models to propagate edits throughout codebases by tracing call graphs---a commonly-studied graph traversal ability \citep{wang2023can}. This axis scales the number of functions models must edit $|V_\text{edit}| = n$.

    \item \textbf{Function Search:} Models often search through large codebases to decide which sections require edits. This axis scales the number of functions agents do \underline{not} need to edit $|V_\text{init}| = n$, mirroring Needle in a Haystack evaluations for long-context reasoning \citep{kamradt2023needle}.

    \item \textbf{Runtime Resolution:} Coding tasks may force models to trace increasingly long computations to resolve runtime values \citep[e.g., CRUXEval]{gu2024cruxeval}.
    To test this, we wrap an \inlinecode{if check():} around identical~code blocks and have models only edit blocks that run.
    We scale the size of resolving \inlinecode{check()} via a Collatz-based function \citep{lagarias19853} whose \inlinecode{for} loop runs for $n$ steps (Code Block~\ref{code:collatz}), requiring $n$ steps of reasoning to resolve.

    \item \textbf{Instruction Following:} Users issue long queries when models should make more edits at once---often assessed via instruction-following benchmarks \citep{pyatkin2026generalizing}. We define task size by appending $n$ distinct instructions to our problem statement, requiring models to also add asserts, comments, temporary values, etc.\ (full list in Appendix~\ref{appendix:example_tasks}).
    This axis is unusable with ``Remove Dead Code'' (\cref{subsection:dag}), as deletion is the only valid action.
    
\end{enumerate}

\section{Experimental Setup} \label{subsection:task_generation}

We generate $6,840$ tasks with \model{} across three code types, five tasks (four in instruction following), four axes, and six sizes---with $20$ tasks each. Varying code and task types lets us more rigorously assess task size scaling per axis.
Below, we discuss our tasks, metrics (\cref{subsection:metrics}), and model design (\cref{subsection:agent}).

\subsection{Tasks} \label{subsection:metrics}
\noindent \textbf{Task Grid.} 
For each axis in \cref{subsection:complexity}, we sweep task size parameter $n \in \{5, 10, 25, 50, 100, 200\}$.
Function~traversal and search fix $|V_\text{start}| = 25$ and $|V_\text{edit}| = 25$, respectively.
We set $|V_\text{start}| = |V_\text{edit}| = 10$ for~runtime resolution/instruction following.
We sample $20$ \graph per configuration, with variation from function calls, operations, \inlinecode{check()} inputs for runtime resolution, and instruction~following~rules.

\noindent \textbf{Success Metrics.} 
Our tasks are rule-based transformations, so we can form ground-truth programs \true.
We then compare model-edited programs \pred with \true via black-box tests \citep{myers2004art}: we verify each function in \pred returns the same outputs as \true's version on $50$ random inputs (i.e., $50|V|$ total tests, so $500$ minimum).
We add rules to verify dead~code~solutions drop all \edit functions, runtime resolution solutions edit valid \inlinecode{if} branches, and instruction following solutions execute all requests.
A model gets a score of $1.0$ per task iff all checks~pass.

\subsection{A Flock of Language Models and Agents} \label{subsection:agent}

We compare how task size affects LLMs and~agents.
The key difference is that~while all~models use natural~language reasoning, agents have tools. 
Parameters are default when unspecified.

\noindent \textbf{Language Models.}
We use GPT \citep[5.4 Mini, 5.3 Codex, 5, and 5.5]{openai_gpt5_system_card_2025},~Grok-4.1 Fast \citep{grok_system_card_2025}, Deepseek-V3.2 \citep{liu2025deepseek}, GPT-OSS 120B \citep{agarwal2025gpt}, and Mistral Large 3.1 \citep{introducing_mistral}.
We adapt SWEAgent's prompt \citep{yang2024sweagent}: LLMs view the problem statement, then return a \texttt{<thought></thought>} block with natural language reasoning followed by a full rewrite of the code  (Prompt~\ref{prompt:llm_full}).
This surpasses generating~diff patches (Appendix~\ref{appendix:model}).
We use medium reasoning effort when~adjustable.

\noindent \textbf{Agents.} We use GitHub Copilot's public harness \citep[\texttt{v1.0.76}]{microsoft_vscode_copilot_chat} with GPT~5.5 and 5.4 Mini, \citep{openai_gpt55_system_card_2025}, Gemini-3.5 Flash and 3.1 Pro \citep{google_gemini3_2025}, Sonnet-4.6,~and Opus-4.7 \citep{anthropic2025claude_sonnet_4_5}. 
We prompt with~the problem statement and~give read-write access to a folder with the input code.
Copilot~uses a three-stage~ReACT loop~\citep{yao2023react}: at each turn, it outputs reasoning $r_i$, a tool call $t_i$ (\inlinecode{bash}), and adds the tool~output~$o_i$.
Agents are costly (Table~\ref{table:cost}, $25$k USD in final runs), so we use just~one~harness.
After~our runs, OpenAI disclosed an incident where agents communicated across environments \citep{openai2026huggingface}.
This spurred us to look for such behavior in Copilot: $11$\% of traces~accessed~shared directories and $<0.2\%$ queried other agents' session history.
However, this minimally~impacts results, as fully sandboxed runs have the same accuracy and tool call patterns (Appendix~\ref{appendix:sandbox}).

\section{Agents Spend Effort Understanding Code, not Just Editing} \label{section:results}

We find that large \model{} tasks challenge LLMs but not agents due~to~tool use (\cref{subsection:compare_agent_llm}).
Agent traces show they spend many tokens on tools to understand code in larger tasks---not just edit it (\cref{subsection:reasoning_trace})---with similar trends in SWE-Bench (\cref{subsection:swe_bench}).
We thus argue that scaling understanding effort is promising for making hard agentic coding tasks, which we test~in~\cref{section:understanding}.

\begin{figure}
    \centering
    \includegraphics[width=\linewidth]{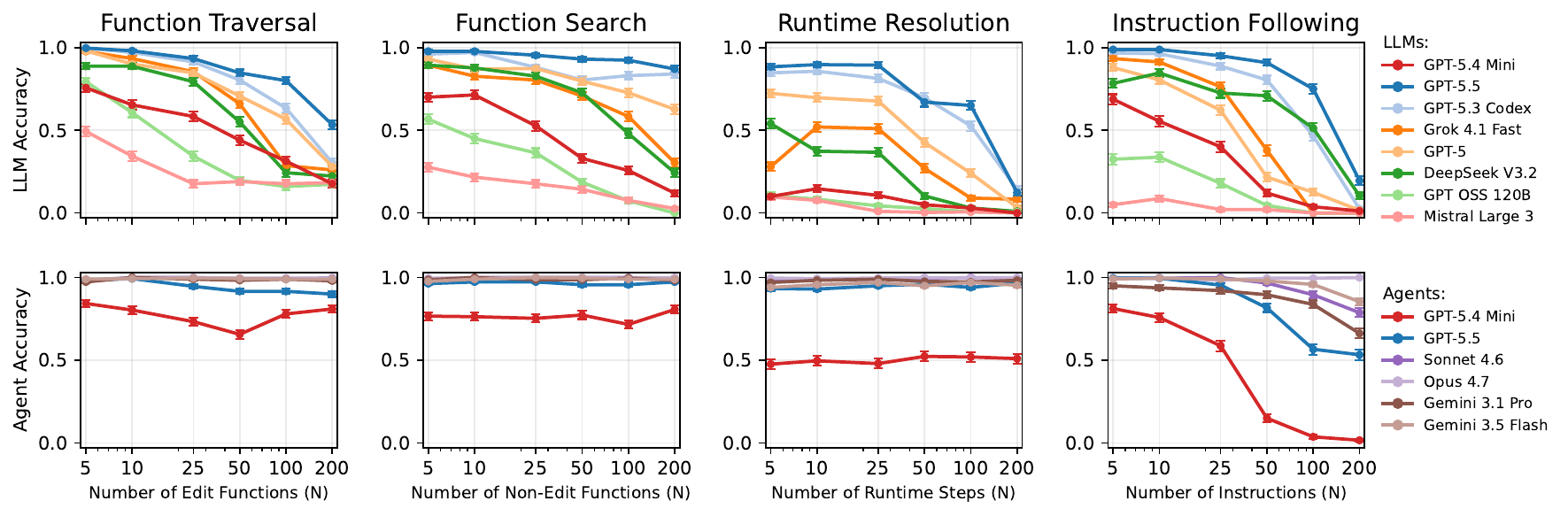}
    \vspace{-4ex}
    \caption{Comparison of LLM and agent accuracy over our four scaling axes, averaged over all tasks with stderr. While all LLMs lower in accuracy as task size $n$ grows, agents often remain near-perfect.
    Appendix~\ref{appendix:task_breakdown} has similar trends when segmenting by each task.
    }
    \vspace{-1ex}
    \label{fig:llm_vs_agent}
\end{figure}

\subsection{Comparison of LLM and Coding Agent Accuracy} \label{subsection:compare_agent_llm}

We display aggregated results on our four complexity axes in Figure~\ref{fig:llm_vs_agent}, with takeaways below:

\textbf{LLM-Hard $\nRightarrow$ Agent-Hard.} LLM accuracy drops with $n$, but agent accuracy is mostly constant and often near perfect.
This is not just due to differences in backbone~LLMs, as GPT-5.4 Mini and GPT-5.5 degrade in accuracy as LLMs, but score consistently as agents (except instruction following).
Appendix~\ref{appendix:task_breakdown} has the same trend with a subset of \model{} tasks with Gemini models.
Tasks of certain complexity or size that challenge LLMs may no longer challenge agents, so evaluating agents' limits will either require scaling task sizes to colossal $n$ \citep{sinha2026the} or new axes for complexity.
We pursue the latter path~in~\cref{section:understanding}.

\textbf{Weaker LLMs can Climb via Harnesses.} GPT-5.4 Mini, when used in an agent harness, surpasses stronger LLMs without tools (GPT-5.5, Grok) in non-instruction following tasks for $n > 100$.
Thus, developers do not always need large LLMs in harnesses; small LLMs that generate and compose tool calls well can generalize to larger tasks \citep{zhang2026harnesses}.

\textbf{Tools Let Agents Graze Codebases.} To explain agents' near-perfect accuracy, we analyze execution traces.
Agents often write and run ``meta-scripts'': \inlinecode{ast} code to move variables in function traversal, \inlinecode{grep} to locate names in function search, or \inlinecode{exec()} to learn outputs in runtime resolution.
LLMs often generate more natural language reasoning as task size grows (Figure~\ref{fig:llm_reasoning_tokens}), but some tasks let agents write scripts with fixed reasoning invariant of~$n$.

The exception is instruction following---rules struggle to map natural language to edits \citep{zan-etal-2023-large}---which is also the only axis where agents degrade ($n=200$, except Opus).~A~new consideration for synthetic tasks with agents is whether tools bypass intended task size.~As we show in \cref{section:understanding}, there are scalable axes beyond instruction following that are hard to~script.

\begin{table}[!h]
\centering
\caption{Our surfaced tool call taxonomy and the distribution of tool use in our evaluations.}
\vspace{-1.5ex}
\label{table:tool_call}
\small
\begin{tabularx}{\linewidth}{@{}
    >{\raggedright\arraybackslash}p{0.17\linewidth}
    >{\raggedright\arraybackslash}p{0.25\linewidth}
    >{\raggedright\arraybackslash}X
@{}}
\toprule
\textbf{Label} & \textbf{Example} & \textbf{Description} \\
\midrule

\textsc{Read (22\%)}
& \inlinecode{view solution.py}
& Reading files or lines of code verbatim. \\

\textsc{Analyze (14\%)}
& \inlinecode{ast.parse(src)}
& Analyzing dependencies, call graphs, runtime, etc. \\

\textsc{Search (11\%)}
& \inlinecode{grep "def check"}
& Pattern matching a parameter, function, or string. \\

\textsc{Edit (36\%)}
& \inlinecode{edit solution.py}
& Changing lines of code in the solution program. \\

\textsc{Test (8\%)}
& \inlinecode{pytest test.py}
& Validating code via syntax checks or test cases. \\

\textsc{Other (9\%)}
& \inlinecode{pip install numpy}
& Any other tool (summary, file cleanup, etc.) \\
\bottomrule
\end{tabularx}
\vspace{-1ex}
\end{table}

\begin{figure}
    \centering
    \includegraphics[width=\linewidth]{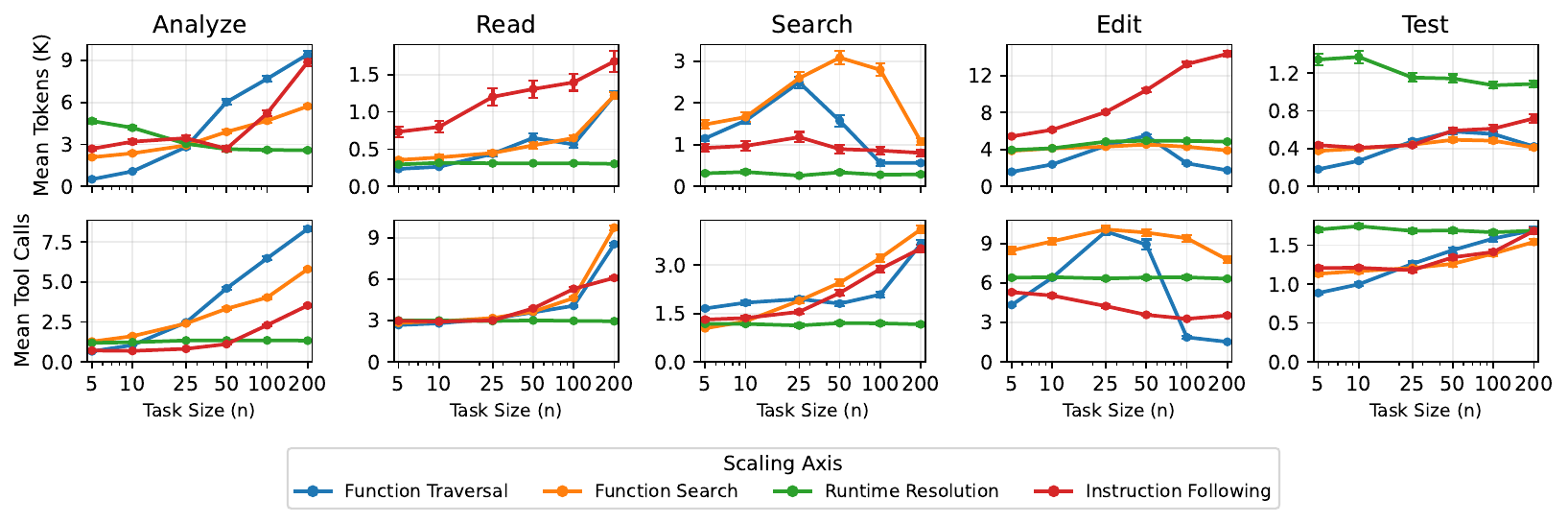}
    \vspace{-4.7ex}
    \caption{Agents' mean tool call count and output tokens for Table~\ref{table:tool_call}'s categories (with~stderr).
    Only understanding tools (\textsc{Read}, \textsc{Analyze}) scale in calls and tokens with task~size in 3/4 axes. \textsc{Edit} only scales in instruction-following. Per-agent results are similar (Appendix~\ref{appendix:tool_breakdown}).}
    \label{fig:cabra_trace}
    \vspace{-1.5ex}
\end{figure}

\subsection{Trace Analysis: Agents Often Spend Tool Calls on Understanding Code} \label{subsection:reasoning_trace} 

We analyze tool use further via a qualitative coding protocol \citep{bingham2023data, balepur-etal-2026-test}: we review a subset of agent~traces and derive six tool types (Table~\ref{table:tool_call}).
We then prompt Qwen3-4B Instruct \citep{yang2025qwen3} to label each tool call in agent traces from \cref{subsection:compare_agent_llm} via its name, inputs, and reasoning tokens---reaching 94\% agreement with 20 author labels per class (details in Appendix~\ref{appendix:classifier}).

For each tool type, we find the number of calls and output tokens used, averaged over agents.
For tool call $t_i$, the latter is the sum of the tokens used to create $t_i$ and tokens in reasoning $r_i$: chain-of-thought between the output $o_{i-1}$ of $t_{i-1}$ and $t_i$ (or after the prompt for $i=0$). 
Token use~is our proxy for effort devoted to each tool type.
Some tokens may link to earlier calls, but token attribution is hard, and \citet{chen2026depo} study step-wise token use like~this.

While prior work uses edits to predict coding task difficulty \citep{deng2026swebench, raghavendra2026swe}, only instruction following elicits more \textsc{Edit} tokens with $n$ (Figure~\ref{fig:cabra_trace}).
Instead, \textsc{Read} and \textsc{Analyze} tokens grow for all axes but~runtime.
We treat these two~as \textbf{code~understanding} actions: acquiring knowledge of a program's construction and reasoning about its behavior to guide later edits \citep{littman1987mental, maalej2014comprehension}.~Their~position~in agent traces backs our interpretation: 48\% of \textsc{Read} and 40\% of \textsc{Analyze} calls occur in the first quarter, and 69\% and 56\%, respectively, precede the first \textsc{Edit}---the most of all tools.
\textsc{Search} may support understanding, but work treats pattern-matching lookups~(\inlinecode{grep}) as distinct actions preceding understanding \citep{wilde2001case, bennett2002software}.~Consistent~with this idea, only 12\% of \textsc{Search} tools are followed by \textsc{Edit}, but 19\% are~followed~by~\textsc{Read}.

Comparing tool use on our scaling axes with agent accuracy on said axes (Figure~\ref{fig:llm_vs_agent}) shows agents excel on tasks where they \textsc{Understand} (\textsc{Read} + \textsc{Analyze}) more (search/traversal), but start to fail if they \textsc{Understand} \textit{and} \textsc{Edit} more (instruction following).
Appendix~\ref{appendix:edit_without_understanding} has~tasks~where \textsc{Edit} use grows more than \textsc{Understand}, validating them as unique~constructs.~This spurs us to ask if tasks that demand \textsc{Understand} tools can thwart agents~in~\cref{section:understanding}.

\begin{figure}[h]

    \centering

    \begin{subfigure}[t]{0.65\columnwidth}
        \vspace{0pt}
        \centering
        \caption{Trace metrics grouped by human-annotated difficulty levels.}
        \vspace{3pt}
        \includegraphics[width=\linewidth]{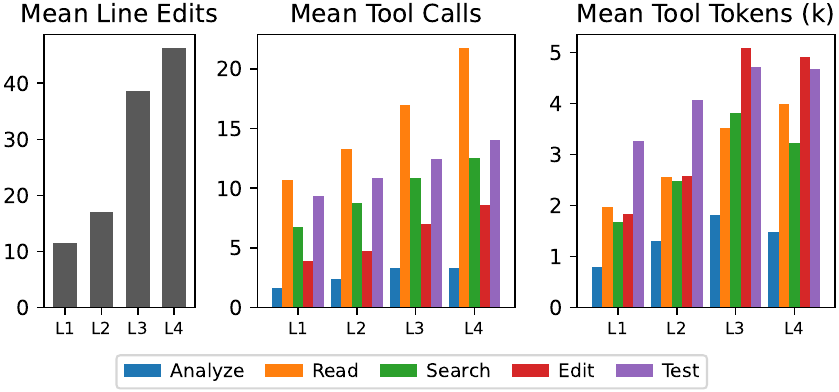}
        \label{fig:breakdown}
    \end{subfigure}\hfill%
    \begin{subfigure}[t]{0.31\columnwidth}
        \centering
        \small
        \caption{Top-eight ranked metrics by agent accuracy correlation.}
        \vspace{5pt}
        \begin{tabular}{lc}
\toprule
Metric & $r\;(\uparrow)$ \\
\midrule
$\textsc{Understand}_\text{num}$ & -0.200 \\
$\textsc{All Tool}_\text{tok}$ & -0.178 \\
$\textsc{Edit}_\text{tok}$ & -0.177 \\
$\textsc{Read}_\text{num}$ & -0.171 \\
Lines Edited & -0.159 \\
$\textsc{Analyze}_\text{num}$ & -0.158 \\
$\textsc{All Tool}_\text{num}$ & -0.157 \\
$\textsc{Edit}_\text{num}$ & -0.151 \\
\bottomrule
\end{tabular}

        \label{fig:correlation}
    \end{subfigure}
    \vspace{-1.4em}
    \caption{Trace analysis on SWE-Bench Verified \citep{jimenez2024swebench} reveals: (a) human-labeled difficulty correlates with edit size and tool use, making failures hard to disentangle;~and (b) \textsc{Understand} tool call count (\textsc{Read} $+$ \textsc{Analyze}) is most strongly linked to agent failures.}
    \label{fig:swe_bench}
    \vspace{-1.5ex}
\end{figure} 

\subsection{Understanding Persists in SWE-Bench Verified}
\label{subsection:swe_bench}

\model{}'s synthetic tasks offer control but lack ecological validity \citep{de2020towards}, so we test our hypothesis---code understanding beyond editing explains task difficulty in agents---on more realistic tasks.
We collect released \texttt{mini-swe-agent} trajectories on SWE-bench Verified \citep{yang2024sweagent} from Claude-4.5 Opus/Sonnet, Gemini-3 Pro/Flash, GPT-5.2, and GPT-5 Mini.
We study how the edited lines of code in the agents' final solution (a commonly-used difficulty metric \citep{deng2026swebench, raghavendra2026swe}) versus our tool call metrics from \cref{subsection:reasoning_trace} (by running our classifier on these traces) relate to two measures of task difficulty: SWE-bench's human-annotated difficulty levels and agents' accuracy (linked to their traces).

\textbf{Edits are not the scapegoat for human-rated difficulty.} SWE-bench tasks come with difficulty levels based on the time developers~estimate humans would need to finish them (L1 $\leq$15 min; L2 = 15 min -- 1 hr; L3 = 1--4 hours; L4 $\geq$4 hours).
Most metrics rise with~level (Figure~\ref{fig:swe_bench}, left):~longer tasks elicit not just more edits, but also reading, analysis, and testing.
Since metrics grow jointly, SWE-bench alone cannot discern when errors stem from edit volume or understanding how to edit---motivating controlled evaluations such~as~\model{}.

\textbf{Understanding predicts errors better than edit count.} We measure the point-biserial correlation $r$ \citep{tate1954pointbiserial} between agent accuracy and each metric.
All $r$ are low: it is tough to predict errors, further motivating \model{}-style tasks for more control.
The number of understanding tool calls (defined as \textsc{Read} plus \textsc{Analyze} calls) is most strongly correlated with accuracy (Figure~\ref{fig:swe_bench}, right), more so than \textsc{Edit} token use and lines edited.
Overall, this further underlines the key role code understanding plays in the evaluation of coding agents.

\section{Intense Understanding Tasks Can Make Coding Agents Bleat} \label{section:understanding}

While work often scales \emph{edit size} toward harder coding agent tasks, \cref{section:results} suggests to~control agent accuracy via an analogous \emph{understanding size}: how much understanding~occurs before editing.
To test this, we apply \model{}'s two-stage evaluation approach:
we~design a synthetic task where models compare lines~of~code across classes to extract shared logic (\cref{subsection:understanding_task}),
then show agent accuracy drops over three scaling axes that target understanding effort (\cref{subsection:understanding_results}).

\begin{figure}
    \centering
    \includegraphics[width=\linewidth]{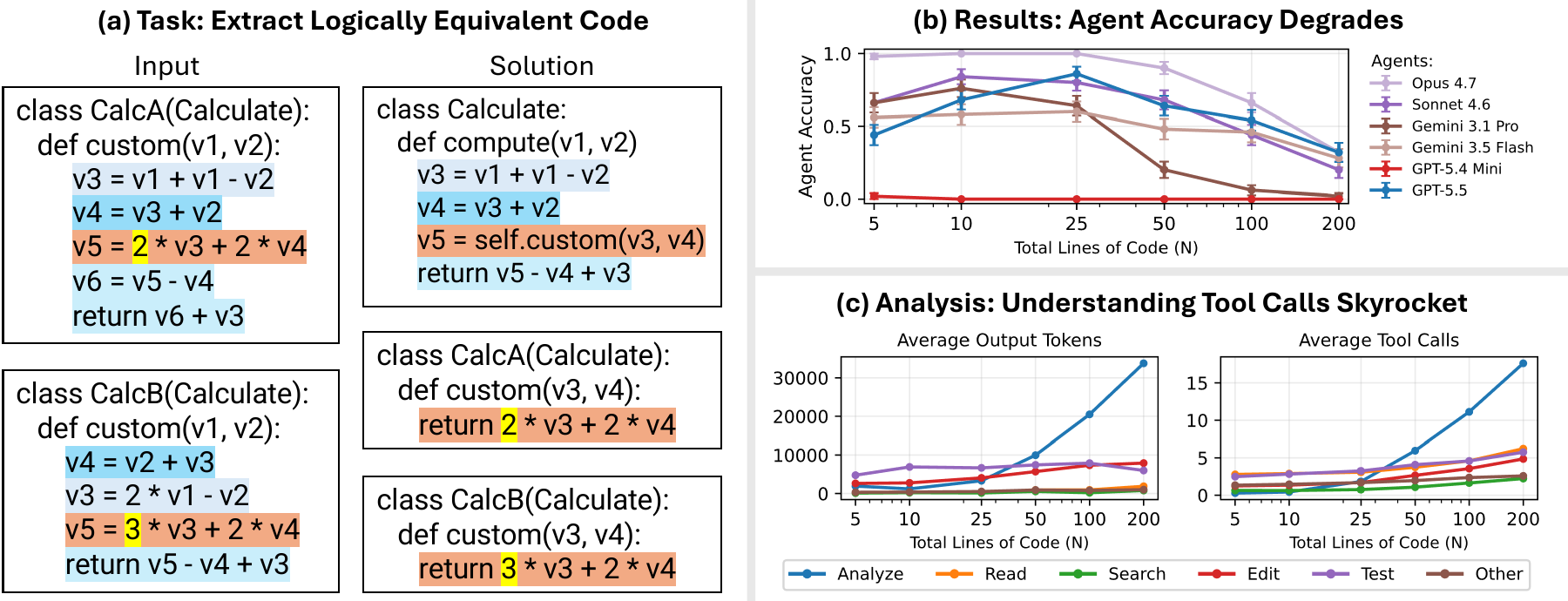}
    \vspace{-4ex}
    \caption{We extend \model{} to an intense understanding task: given two classes with shared logic, move \hlblueone{logically }\hlbluetwo{equivalent }\hlbluethree{code} into a parent class while preserving \hlred{per-class differences} (a).
    Agent accuracy plummets as lines of code grow (b) and understanding tools balloon (c).}
    \label{fig:merge_code}
    \vspace{-1ex}
\end{figure}

\subsection{A Code Understanding Task: Extracting Logically Equivalent Code} \label{subsection:understanding_task}

We extend \model{} to a code understanding task based on program equivalence \citep{pitts1997operationally}, which \citet{wei2025equibench} use to test if models understand code semantics.
We define~calculator classes as DAGs $\mathcal{C}=\langle V,E\rangle$, where $v_i \in V$ is a line defining a variable and edge $(v_i,v_j) \in E$ means $v_j$ uses $v_i$ in its computation (DAG visualized in Figure~\ref{fig:merge_codebases_dag}).
Models~start with classes $\mathcal{C}_A$, $\mathcal{C}_B$ with $\ell$ lines each that differ in behavior at $d$ lines: models must move the~$\ell-d$ equivalent lines into a parent class $\mathcal{C}_\text{parent}$ and keep the $d$ differences in child classes (Figure~\ref{fig:merge_code}, left).
As $\ell$ grows, edits scale modestly (e.g., $O(\ell)$ lines per class), but combinations of lines that can be understood to locate differences skyrockets (e.g., $\Omega(\ell^2)$ lines worst case).
We later validate this intuition empirically: agents' tool calls for \textsc{Analyze} rise sharply with~$\ell$~(\cref{subsection:understanding_results}).

We instantiate tasks by first sampling $\ell$ line nodes; each line defines a~new variable~by chaining input parameters, past variables, and non-zero integer coefficients in $[-5, 5]$ via sums and products.
We randomly select $d$ divergent line nodes, done by either: 1) flipping a sign; 2) editing a coefficient; 3) editing a variable; or 4) adding an extra term.
To make the code surrounding these differences behaviorally equivalent while syntactically dissimilar, we use five transformations: 1) rename variables; 2) shuffle commutative lines; 3) decompose/merge lines of code; 4) distribute multiplication ($3a = a + a + a$); and 5) swap term order ($a + b = b + a$).

We prompt agents with one example and ask for the most efficient edit, i.e., keep as much code as possible in the parent class (Prompt~\ref{prompt:merge_code}).
Agents receive a score of $1$ iff they~preserve behavior (black-box testing, \cref{subsection:metrics}) and each child class has $d$ different lines of code maximum.

\begin{figure}[t]
    \centering
    \begin{subfigure}[t]{0.68\columnwidth}
        \centering
        \caption{Mean agent accuracy after adding/removing transforms ($n=50$).}
        \vspace{-1pt}
        \includegraphics[width=\linewidth]{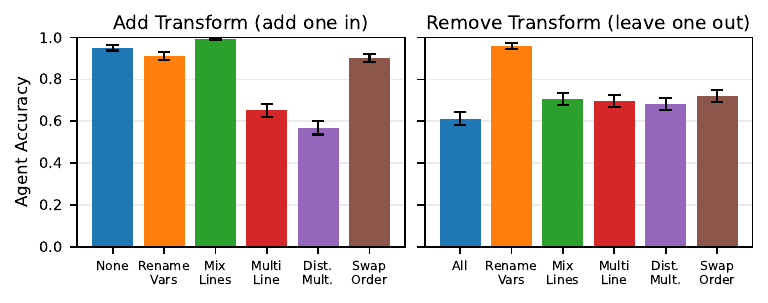}
        \label{fig:transform_equal}
    \end{subfigure}\hfill%
    \begin{subfigure}[t]{0.28\columnwidth}
        \centering
        \caption{Mean agent accuracy with varied difference count.}
        \vspace{-1pt}
        \includegraphics[width=\linewidth]{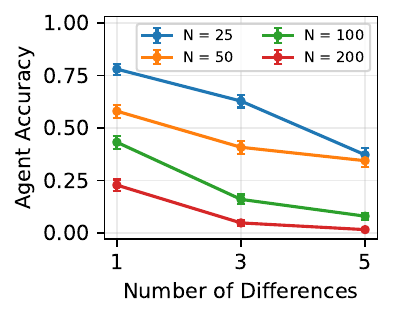}
        \label{fig:num_differences}
    \end{subfigure}\hfill%
    \vspace{-1.5em}
    \caption{We reveal agent difficulties in extracting equivalent code: aligning identical~decompositions (left), mapping renamed variables~(middle), and locating differing lines (right).}
    \vspace{-1ex}
    \label{fig:ablation}
\end{figure}

\subsection{Understanding When Understanding is Challenging} \label{subsection:understanding_results}

We now scale the size of our code equivalence task (\cref{subsection:understanding_task}) to probe agents' code~understanding abilities.
We first confirm that scaling one axis---(1) the total lines of code per class---raises understanding effort and lowers accuracy.
We then study two more scaling axes: (2) the~total transformations agents reason over; and (3) the number of differing lines they must~locate.

\textbf{1) Comparing Lines of Code.}
We sample 50 tasks while scaling $\ell \in \{5, 10, 25, 50, 100, 200\}$, increasing the combinations of lines of code that could be understood and compared. 
Higher~$\ell$ drops agent accuracy (Figure~\ref{fig:merge_code}b),\footnote{Prompted LLMs from \cref{subsection:agent} are much less accurate (Appendix~\ref{appendix:merge}).} and elicits far more \textsc{Understand} than \textsc{Edit} tools~(Figure~\ref{fig:merge_code}c)---supporting that understanding is a weakness in coding agents that can be~improved.

\textbf{2) Aligning Logic.} We hypothesize that our understanding task is hard as it~requires~aligning syntactically distinct but semantically equivalent code \citep{10.1145/3314221.3314596}.
Thus,~we test the equivalence transform types agents must align as complexity axes for understanding.
We ablate transforms at~$\ell=50$: adding each to a no-transform baseline (Figure~\ref{fig:ablation},~left) and removing each from an all-transform baseline (Figure~\ref{fig:ablation}, middle).
Multi-line~edits~and~distributive multiplication~cause~large drops when added---for Gemini specifically (Appendix~\ref{appendix:merge})---failing on multi-hop alignment \citep{zhang-etal-2025-belle}.
Removing variable~renaming boosts accuracy most, so agents may rely on~shallow naming to understand~code \citep{lam2025codecrash}.

\textbf{3) Locating Differences.} When agents edit, they generally follow the two-step process of: (i) locating lines that differ; and (ii) executing the refactor.
To disentangle where errors occur, we study both metrics in \cref{subsection:understanding_task}.
Agents preserve code behavior in over 90\% of cases, so~most failures stem from (i) versus (ii); agents keep too many lines in the child class, failing to find one line differences.
We further probe this weakness by varying the number of differing lines $d \in \{1, 3, 5\}$ while holding lines of code $\ell$ fixed.
Accuracy erodes as $d$ grows (Figure~\ref{fig:ablation}, right), validating agents struggle increasingly as they need to identify more behavioral differences.

\section{Related Work}
\label{section:related_work}

To ground \model{}, we synthesize work on code evaluations, agent errors, and synthetic tasks.

\textbf{Evaluating Code Models.} Code generation is a long-studied task \citep{hindle2016naturalness}.
Early evaluations used NL-to-code translation tasks \citep{oda2015ase:pseudogen1, yin2018mining}, then scaled difficulty with algorithm tasks based on contests like LeetCode \citep{chen2021codex, li2022competition, jain2025livecodebench}.
The boom of agents that compose LLMs and tools \citep{yao2023react} motivated harder tasks like issue resolution \citep{jimenez2024swebench, zan2026multi} and full app generation from specifications \citep{si-etal-2025-design2code, zhao2025commit}.
To scale difficulty, work adds more edits \citep{deng2026swebench, raghavendra2026swe}, multi-turn prompts \citep{le2025swe, Laban2026LLMsCY}, and extra requirements \citep{ding-etal-2026-octobench, yang2026programbench}.
We also show that understanding is a less studied but key axis to control~difficulty.

While such tasks simulate real use, they struggle to diagnose model~errors, as they~differ~on many confounded axes.
We review these limitations then discuss synthetic tasks as a solution.

\refstepcounter{subsection}
\label{subsection:rw_agent_failures}
\textbf{Agent Failures.} Knowing model failures drives progress.
For coding agents, work often studies properties of failed traces on repository benchmarks \citep{cemri2026why} like~execution strategy, tool call type, and length---relating them to~agent accuracy to guide methods~\citep{Majgaonkar2025UnderstandingCA, Liu2025AnES, mehtiyev2026beyond}.
However, analyses are correlational---tasks also vary in codebase and task---making it hard to attribute failures, and correlations are often poor (\cref{subsection:swe_bench}).
We argue \model{}-style tasks that control such confounders are better for this goal, which we use to surface agent weaknesses in code understanding.

\refstepcounter{subsection}
\label{subsection:rw_task_complexity}
\textbf{Synthetic Tasks.} Compared to repository benchmarks, synthetic tasks can better probe model~errors \citep{merrill-etal-2020-formal}.
Tasks like bAbI \citep{weston2016babi, tamari-etal-2022-dyna} and Needle in a Haystack \citep{hsieh2024ruler, kim2025one} test limits of a model ability (e.g., long-context) by defining tasks (e.g., locate a passkey) and scaling their size (e.g., corpus size); we use this two-stage approach for \model{}.
Recent work extends~this~to reasoning LLMs: LongCoT for long-horizon execution \citep{motwani2026longcot} and~RELIC for context-free grammar recognition \citep{petty2026relic}.
In code specifically, related work includes CruxEval for code tracing \citep{gu2024cruxeval} and Bug in the Code Stack for codebase search \citep{lee2024bug}, which form our initial \model{}~tasks.
There are also synthetic RL environments to train coding agents \citep{yang2026swesmith, zhu2025trainingversatilecodingagents}, but their use of existing repositories or LLM-generated data impede full control over tasks, unlike \model{}.

Different from all of these works, we explore task complexity in coding agents.
Our main~contribution is exposing that many ``LLM-hard'' tasks are trivial with tool use---task complexity must evolve for agents---then surfacing and probing code understanding as a model weakness.

\section{Conclusion: Raising the Baa-r for Coding Agent Evaluations} \label{section:conclusion}

Using \model{}, we show the number of edits alone is a poor proxy for coding agent task difficulty and instead surface code understanding as a key factor. Across our initial task axes, editing effort scales insufficiently with task size. In contrast, we reveal that code understanding~more often scales~with task size across diverse axes and more reliably predicts SWE-bench Verified errors.
We establish this link concretely with our code merging~task and surface facets like multi-line decomposition and readability where agents still struggle, motivating threads for future~work.

\textbf{Controlled Evaluation.} \model{}-style tasks can both attribute agent error and form axes to probe them.
We use tools for diagnosis, but future work can study problem-solving methods \citep{gao2024strategyllm} and error recovery \citep{huang2023large}.
The next step is imbuing such weaknesses into SWE-bench-style tasks as more realistic evaluations \citep{saxon2024benchmarks}.

\textbf{Agent-Understandable Code.} Agents struggle to compare programs with renamed~variables (\cref{subsection:understanding_results}), showing some code types are harder to understand.
Future work could test if repository indexes \citep{luo2026closing}, memory \citep{zhang2025survey}, and readability practices \citep{buse2009learning} help agents.
As understanding can encompass up to 70\% of SWE work \citep{minelli2015know}, 19\% of real coding requests \citep{baumann2026swe}, and already demands effort in \model{} and SWE-Bench, progress here is key for maintainability \citep{le2025swe} and limiting context rot \citep{laban2026llms}.

\textbf{A Theory of Agents.} Work on transformers' limits \citep{bhattamishra-etal-2020-ability} often uses synthetic tasks like \model{} to empirically test theories, but tasks that were once hard~for~LLMs can now be trivial with tools (\cref{subsection:compare_agent_llm}).
Discerning ``agent-hard'' or ``-easy''~tasks~will require formalizing the work used to call tools and work tools can execute \citep{meyerson2025position}.

\section*{Acknowledgments}
We thank Jennifer Neville and the Microsoft Research AI Interaction and Learning group for their feedback.
We are also grateful for the discussions with other researchers on earlier drafts of this work, namely Grace Chen, Dang Nguyen, Yu Hou, Dayeon Ki, Vishakh Padmakumar, Sarah Wiegreffe, Nikhil Sharma, Shuhaib Mehri, Saujas Vaduguru, Jordan Boyd-Graber, Rachel Rudinger, and the Pangram team.
Nishant is grateful to the MSR NYC intern cohort for making the summer in New York City memorable, and especially to Juhyun Oh, Katelyn Mei, Jessica Bo, Boyi Wei, Jongho Park, Jack Cai, Tao Long, and Sahana Subramanyam for many hours of non-research activities, la la la!
Finally, Nishant would like to thank La Cabra coffee shop on Lafayette Street for providing a reliable source of caffeine throughout the internship---enough to inspire the name of our \model{} framework.

This material is based upon work supported by the National Science Foundation under Grant No. \abr{dge}-2236417.
Any opinions, findings, and conclusions or recommendations expressed in this material are those of the author(s) and do not necessarily reflect the views of the National Science Foundation.

\section*{AI Use Statement}

In this work, we used generative AI tools primarily for code implementation and data visualization (e.g., plots).
We have not used generative AI for writing from scratch, figure generation, or designing the high-level framing and structure of this paper.
We also minimally used generative AI tools for revising writing, brainstorming, reviewing our paper, and literature search.
For coding, we have reviewed all AI-assisted work through a combination of manually defined test cases, and visualization tools to verify that our evaluations generate data, elicit model responses, and score responses correctly.
Before releasing our code, one author reviewed all generative AI outputs to detect any errors.
We take full responsibility for the final content of this work, including any errors from generative AI.

\section*{Ethics Statement} \label{section:ethics}

Our paper shows that coding agents can struggle to understand code.
Given the importance of understanding in real developer workflows \citep{minelli2015know, baumann2026swe}---coupled with evidence that programmers struggle to understand their code when working with agents \citep{shen2026ai, balepur2026paired}---there are risks for failing to address this problem.
We hope that our paper brings more attention to evaluating understanding in coding agents.

\section*{Reproducibility Statement} \label{section:reproducibility}
In Appendix~\ref{appendix:model}, we provide more details on our model setup to help researchers reproduce our results.
Our code is available at \url{https://github.com/microsoft/CABRA}. Our task data, results, and sandboxed Copilot traces are available at \url{https://huggingface.co/datasets/microsoft/CABRA}. Contact the authors for any further inquiries.

\bibliography{references}
\bibliographystyle{iclr2027_conference}

\appendix

\clearpage

\renewcommand{\thefigure}{\thesection.\arabic{figure}}
\setcounter{figure}{0}

\renewcommand{\thetable}{\thesection.\arabic{table}}
\setcounter{table}{0}

\section{Appendix}

\subsection{DAG Task Sampling Procedure} \label{appendix:sample_tasks}

\begin{figure}
    \centering
    \fbox{\includegraphics[width=\linewidth]{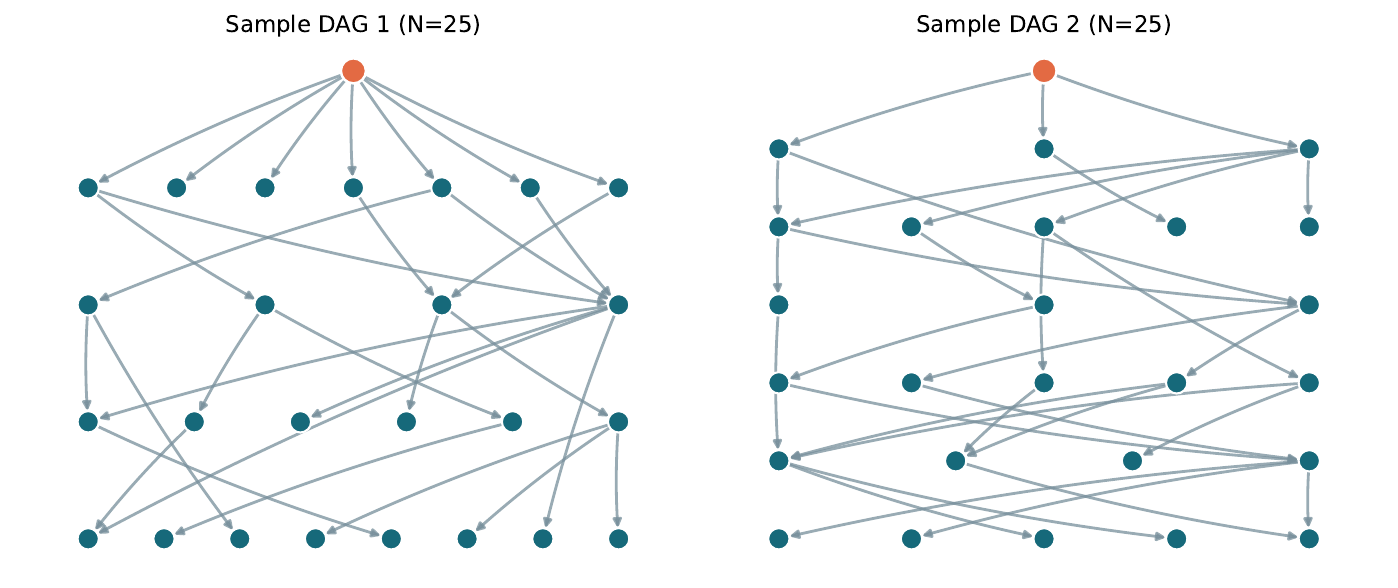}}
    \caption{Examples of $|V|=25$ DAGs sampled from the procedure described in \cref{subsection:dag}.}
    \label{fig:visualize_dag}
\end{figure}

We now describe our sampling procedures for the DAG-based tasks in \cref{section:static}. 
Given $N$ vertices, we sample a height~$h \sim \mathrm{Uniform}(0.45, 0.55)$ and set the topological level count to $\texttt{round}(N^h)$.
We set vertices across levels via Poisson draws, where mean~$\lambda$~is~the number of vertices left to assign divided by the number of levels left.~For each non-root vertex $v_j$, we sample an incoming edge $(v_i, v_j)$ from an earlier level with an offset $\Delta \sim \mathrm{Geom}(0.9)$, biasing edges to adjacent levels but allowing skips (i.e., $i + \Delta = j$). Figure~\ref{fig:visualize_dag} shows~example DAGs and Algorithm~\ref{alg:sample-dag} summarizes the sampling procedure.

\subsection{Example Programs} \label{appendix:example_tasks}

We present examples of math, string, and array programs for the Add Parameter task (\cref{subsection:dag}) in Code Block~\ref{code:math}, \ref{code:string}, and \ref{code:array}, respectively.
Code Block~\ref{code:collatz} shows the Collatz-inspired \inlinecode{check()} function \citep{lagarias19853} we use for runtime resolution, and Code Block \ref{code:calculator} shows an example of two \inlinecode{Calculator} classes in the code merging task (\cref{section:understanding}); the new line separates the synonymous from divergent code, but this formatting is stripped during evaluation. 

For the instruction following task, we use the following bank of rules:
\begin{itemize}
    \item Add a number: \inlinecode{new\_var = var + rand}.
    \item Subtract a number: \inlinecode{new\_var = var - rand}.
    \item Multiply by a number: \inlinecode{new\_var = var * rand}.
    \item Divide by a number: \inlinecode{new\_var = var / rand}.
    \item Add a character prefix: \inlinecode{new\_var = rand + var}.
    \item Add a character suffix: \inlinecode{new\_var = var + rand}.
    \item Replace a character: \inlinecode{new\_var = var.replace(rand, rand2)}.
    \item Surround a string with characters: \inlinecode{new\_var = rand + var + rand2}.
    \item Assert that a value is unequal to a sampled value: \inlinecode{assert var != rand}.
    \item Assert an upper bound: \inlinecode{assert var < rand}.
    \item Assert a lower bound: \inlinecode{assert var > rand}.
    \item Print a value with a prefix: \inlinecode{print(prefix, var)}.
    \item Print a value with a suffix: \inlinecode{print(var, suffix)}.
    \item Add an exact comment of the form \inlinecode{\# label: var}.
\end{itemize}

Sampling values for \inlinecode{rand}, \inlinecode{prefix}, and \inlinecode{label} allows us to scale the total number of unique instructions.

\subsection{Model Implementation Details} \label{appendix:model}

We run all LLMs via litellm\footnote{https://github.com/BerriAI/litellm} using the following models:
\begin{itemize}
        \item \texttt{gpt-5.3-codex\_2026-02-24}
        \item \texttt{gpt-5\_2025-08-07}
        \item \texttt{gpt-5.4-mini\_2026-03-17}
        \item \texttt{gpt-5.5\_2026-04-24}
        \item \texttt{grok-4-1-fast-reasoning\_1}
        \item \texttt{DeepSeek-V3.2\_1}
        \item \texttt{gpt-oss-120b\_1}
        \item \texttt{Mistral-Large-3\_1}
\end{itemize}
        
All models use the Completions API, except Codex (Responses API). We enable reasoning for OpenAI and Grok models---which is \texttt{medium} by default, except for Codex and GPT-5.5---where we set it manually.
We use default parameters otherwise to simulate out-of-the-box behavior \citep{garces-arias-etal-2025-decoding}.
We prompt models until they generate a parseable output codebase---ensuring failures stem from weaknesses in coding abilities rather than output formatting---which typically takes no more than ten tries.

\begin{figure}
    \centering
    \includegraphics[width=\linewidth]{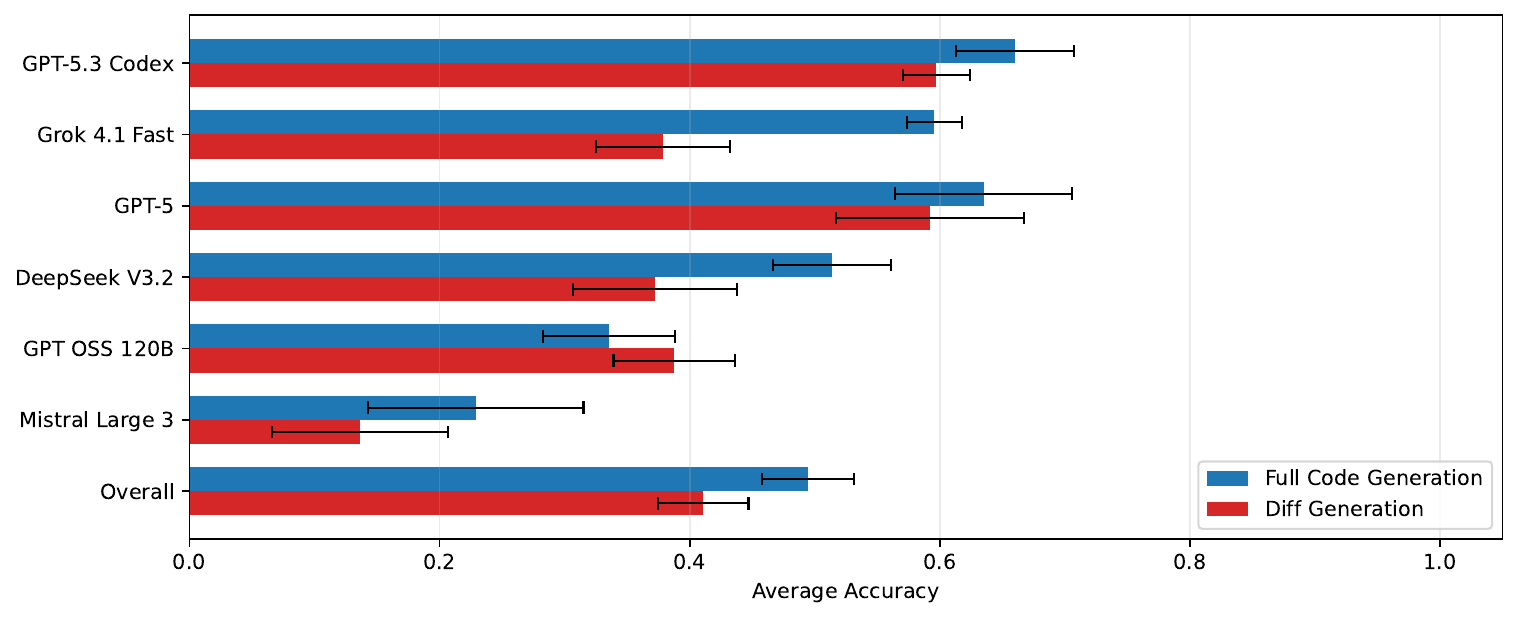}
    \caption{Comparison of LLM task completion when editing code via full codebase regeneration versus structured diff patches when both \start and \edit sweep $n \in \{5, 10, 25, 50, 100, 200\}$. The former yields higher accuracy overall, so we use this format for remaining experiments.} \label{fig:appendix:diff_vs_patch}
    \label{fig:code_vs_diff}
\end{figure}

We evaluate whether LLMs should perform full codebase generation or diff patch generation in \model{} by empirically testing both.
We run a subset of LLMs in \cref{subsection:agent} on a task that combines function traversal and search---scaling \start and \edit with  $n \in \{5, 10, 25, 50, 100, 200\}$.
Full codebase generation yields higher accuracy than patch generation overall, and for all models except GPT OSS 120B (Figure~\ref{fig:appendix:diff_vs_patch}).
Thus, we use it as the configuration in all experiments.
This replicates results of \citet{cheng-etal-2026-diff}, who find structured diff formats can lead to unstable LLM responses.

For the agent harness, we script Copilot \texttt{v1.0.76} (run in July 2026). We script the model using the following command:

\lstdefinestyle{bash}{
  language=bash,
  basicstyle=\ttfamily\small,
  backgroundcolor=\color{bashbg},
  frame=single,
  rulecolor=\color{bashborder},
  framesep=5pt,
  linewidth=\linewidth,
  xleftmargin=0pt,
  xrightmargin=0pt,
  framexleftmargin=4pt,
  framexrightmargin=4pt,
  columns=fullflexible,
  keepspaces=true,
  showstringspaces=false,
  breaklines=true,
  breakatwhitespace=false,
  mathescape=true,
}

\begin{center}
\begin{minipage}{0.96\linewidth}
\begin{lstlisting}[style=bash]
copilot \
  --model $m$ \
  -C $c$ \
  -p $p$ \
  --mode autopilot \
  --silent \
  --allow-all-tools \
  --session-id=$i$ \
  --no-ask-user \
  --no-remote
\end{lstlisting}
\end{minipage}
\end{center}

where $c$ is the path to where we write the \model{} files, $p$ is the problem statement, and $i$ is a randomly-generated UUID so we can access the session later. The other commands remove output streaming, ensure Copilot does not ask clarification questions, give access to all tools, and keep execution local.
The models $m$ are one of the following endpoints:
    \begin{itemize}
        \item \texttt{claude-sonnet-4.6}
        \item \texttt{claude-opus-4.7}
        \item \texttt{gpt-5.5}
        \item \texttt{gpt-5.4-mini}
        \item \texttt{gemini-3.1-pro-preview}
        \item \texttt{gemini-3.5-flash}
    \end{itemize}

We retry until Copilot returns a parseable output codebase, which typically takes no more than three tries at 10 minutes per try (with the exception of the challenging logical equivalence task from \Cref{subsection:understanding_task}, where runtimes between 10 and 30 minutes are common).

\subsection{Classifier Implementation} \label{appendix:classifier}

We classify tool calls using \texttt{Qwen/Qwen3-4B-Instruct-2507} \citep{yang2025qwen3} with $0$ temperature, a context length of $10000$ tokens, and \texttt{sglang} for efficiency.
Along with the six labels in Table~\ref{table:tool_call}, we also classify tools as \texttt{summary} (generates a summary of output changes) and \texttt{plan} (creates a TODO list); we find these are rare, so we merge them with \texttt{other} for analyses.
In rare cases, Qwen output an invalid label; these instances were routed to \texttt{gpt-4.1\_2025-04-14} as a slower but more powerful fallback.
To validate the classifier, we sample $20$ examples for each of the Qwen-predicted labels---yielding $160$ in total.
One author then blindly labels each tool call.
We repeat this process for the initial \model{} tasks (\cref{subsection:reasoning_trace}), SWE-Bench tasks (\cref{subsection:swe_bench}), and merge codebases task (\cref{subsection:understanding_results})---which achieved agreement of 94\%, 89\%, and 94\%, respectively.

\subsection{Further \model{} Accuracy Across Task Types and Models} \label{appendix:task_breakdown}

\begin{figure}
    \centering
    \includegraphics[width=\linewidth]{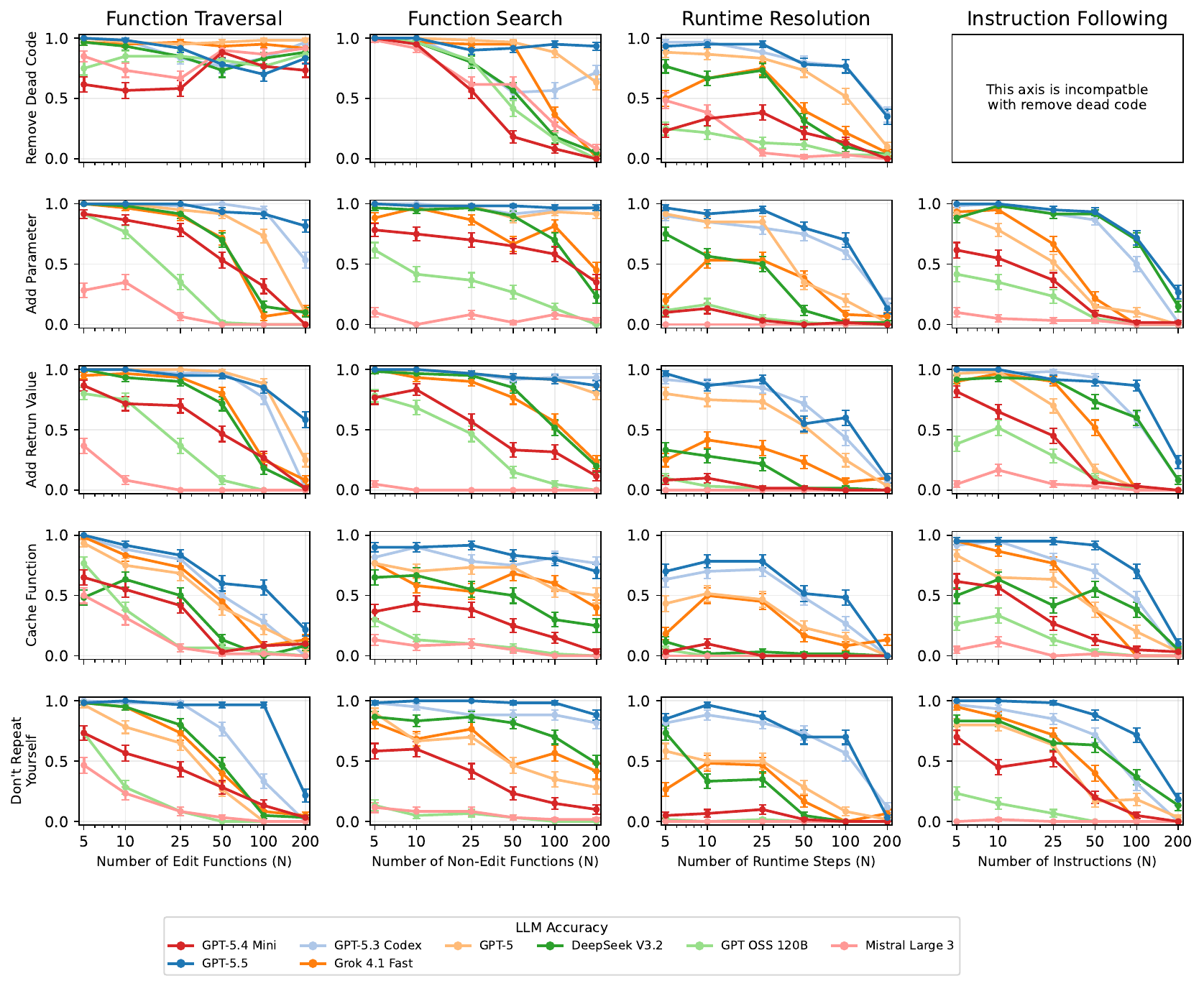}
    \caption{Evaluation of \model{} tasks in Figure~\ref{fig:llm_vs_agent} with \textbf{LLMs}, segmented by the task types in \cref{subsection:dag}. The trend persists: higher complexity tasks tend to challenge LLMs.
    }
    \label{fig:task_llm_breakdown}
\end{figure}

\begin{figure}
    \centering
    \includegraphics[width=\linewidth]{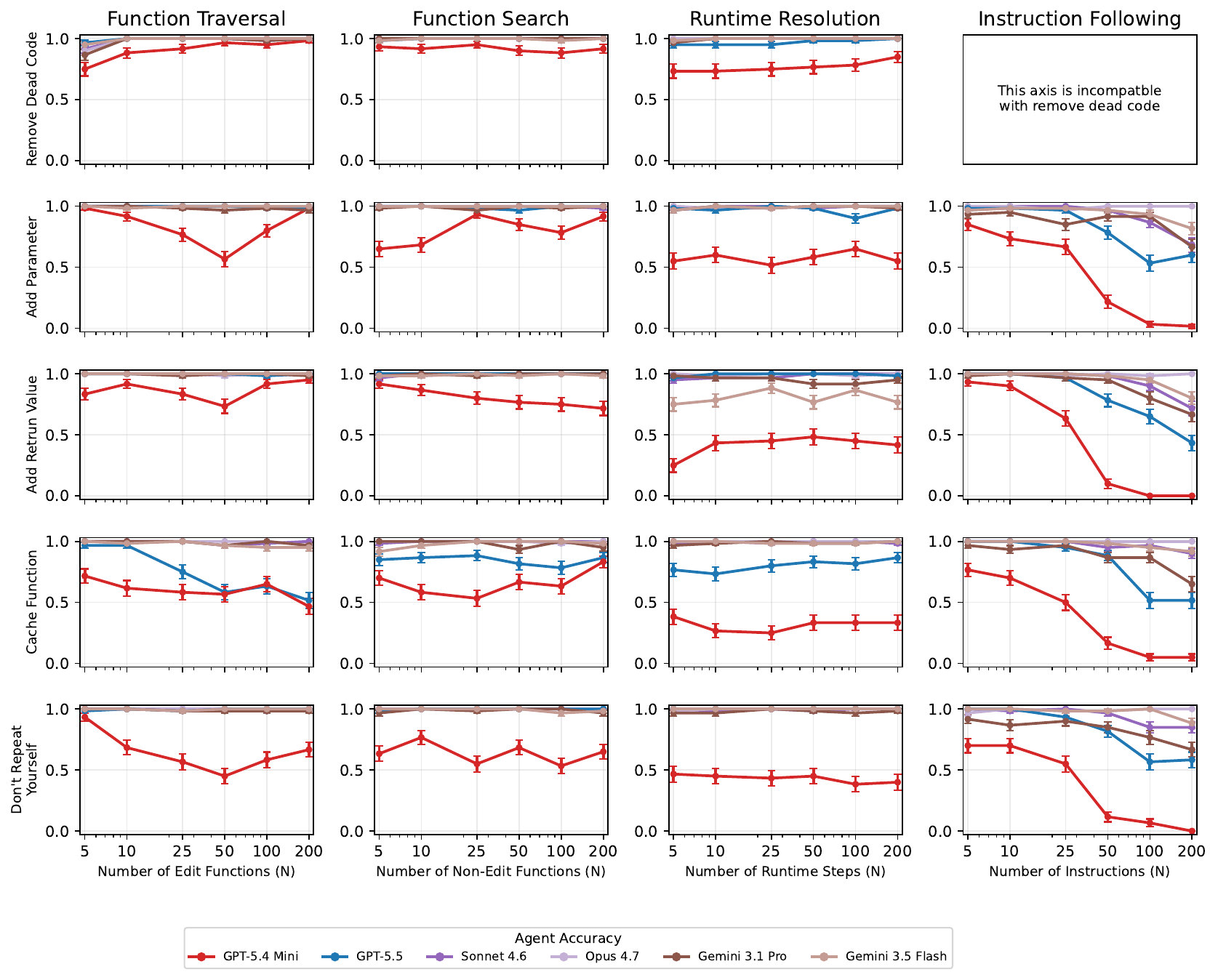}
    \caption{Evaluation of \model{} tasks in Figure~\ref{fig:llm_vs_agent} with \textbf{agents}, segmented by the task types in \cref{subsection:dag}. The trend persists: the strongest agents often have near-perfect scores on larger tasks.
    }
    \label{fig:task_agent_breakdown}
\end{figure}

\begin{figure}
    \centering
    \includegraphics[width=\linewidth]{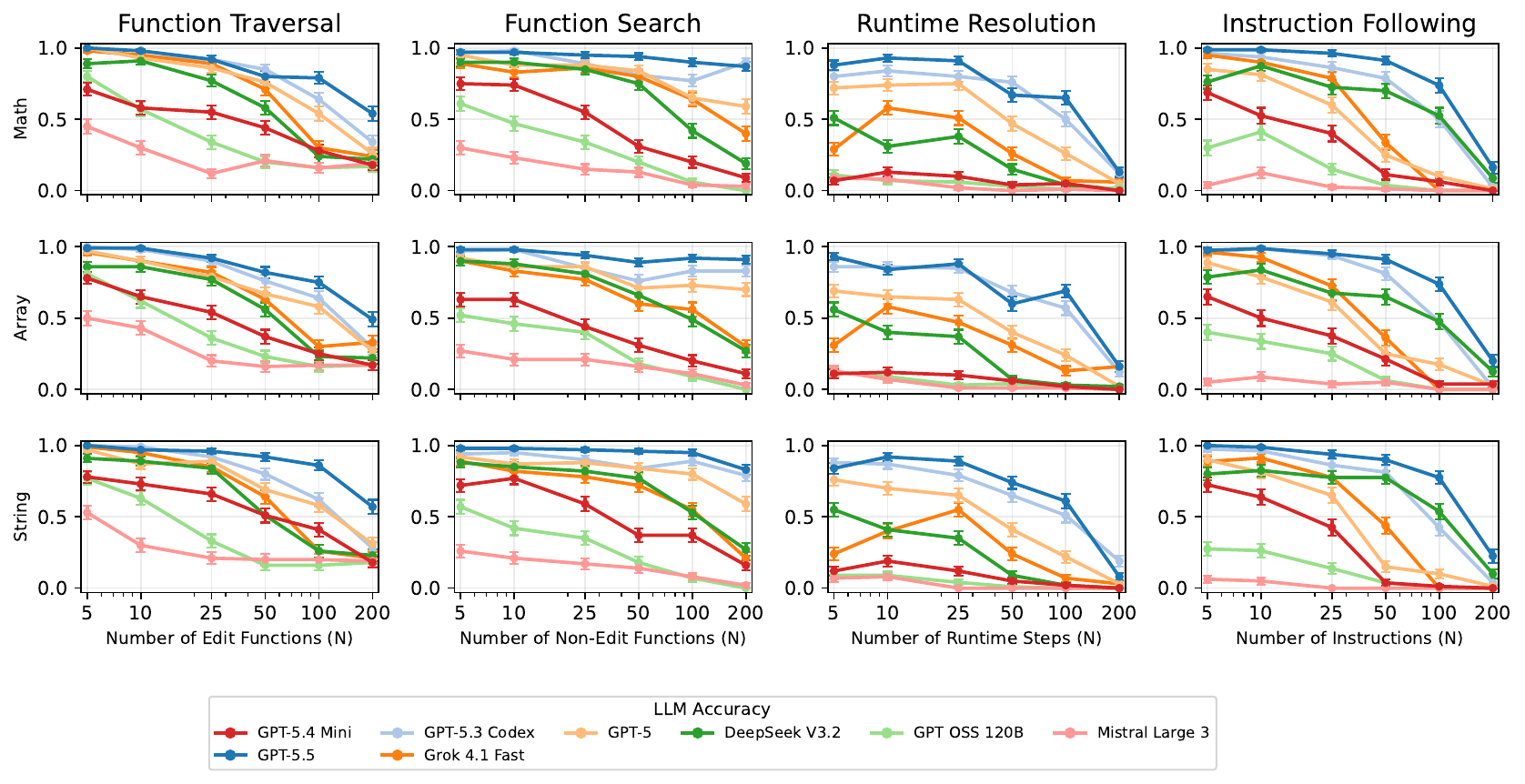}
    \caption{Evaluation of \model{} tasks in Figure~\ref{fig:llm_vs_agent} with \textbf{LLMs}, segmented by the three program types (math, string, array) in \cref{subsection:dag}. The trend persists: higher complexity tasks tend to challenge LLMs.
    }
    \label{fig:program_llm_breakdown}
\end{figure}

\begin{figure}
    \centering
    \includegraphics[width=\linewidth]{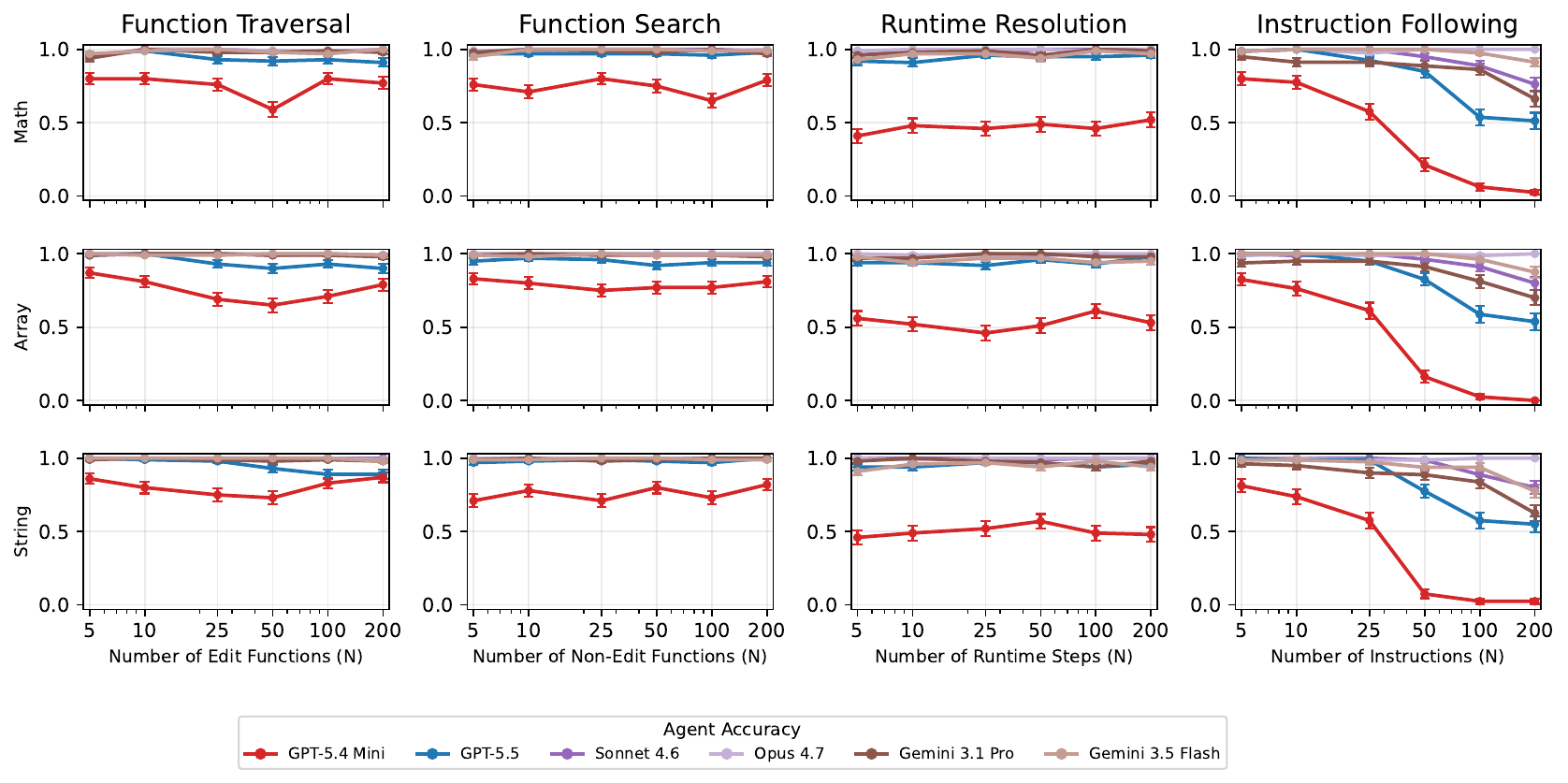}
    \caption{Evaluation of \model{} tasks in Figure~\ref{fig:llm_vs_agent} with \textbf{agents}, segmented by the task types in \cref{subsection:dag}. The trend persists: the strongest agents often have near-perfect scores on larger tasks.
    }
    \label{fig:program_agent_breakdown}
\end{figure}

We decompose results from Figure~\ref{fig:llm_vs_agent} across two dimensions as a robustness check: 1) the five task types; and 2) the three program types (math, string, array; all described in \cref{subsection:dag}).
We present the task breakdowns for (1) across LLMs and agents in Figure~\ref{fig:task_llm_breakdown} and Figure~\ref{fig:task_agent_breakdown}, respectively, and program breakdowns for (2) across LLMs and agents in Figure~\ref{fig:program_llm_breakdown} and Figure~\ref{fig:program_agent_breakdown}, respectively.
Trends persist even in these decompositions: higher-complexity tasks tend to challenge LLMs, but agent accuracy remains nearly perfect.

\begin{figure}
    \centering
    \includegraphics[width=\linewidth]{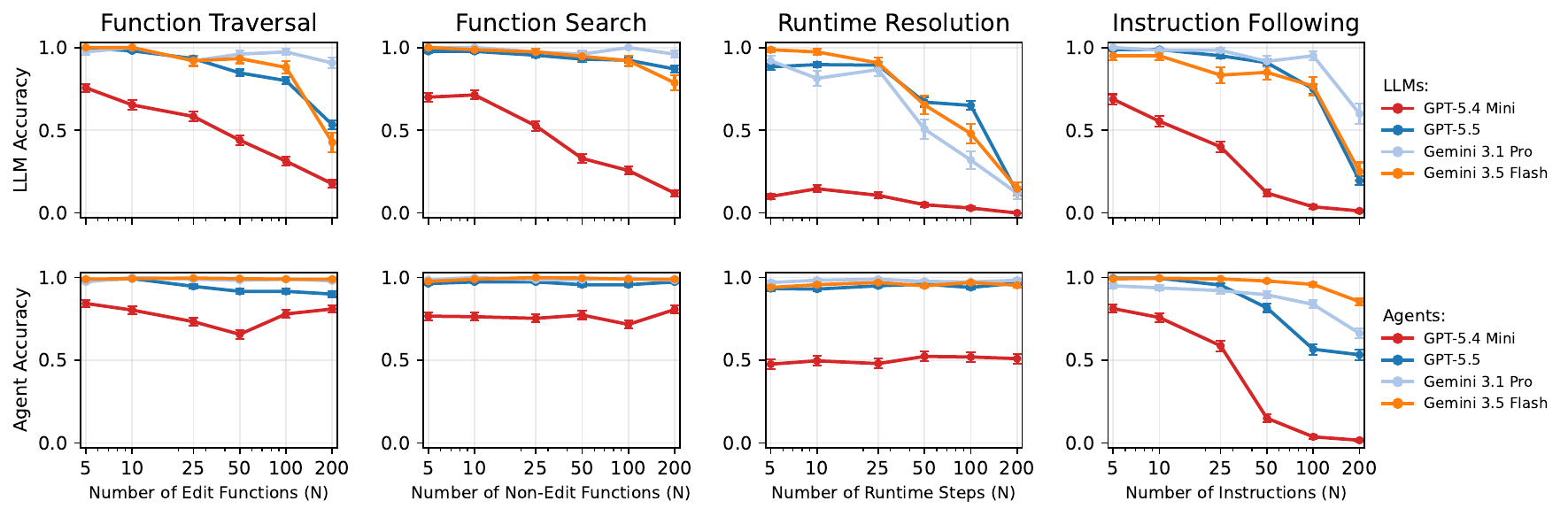}
    \caption{Evaluation of reasoning LLM and agent accuracy when controlling for baseline models: GPT-5.5, GPT-5.4 Mini, Gemini-3.1 Pro, and Gemini-3.5 Flash. The trend is consistent with \cref{subsection:compare_agent_llm}: LLM accuracy degrades more when $N$ scales versus agent accuracy.
    }
    \label{fig:gemini_gpt}
\end{figure}

Finally, we also run Gemini-3.5 Flash and Gemini-3.1 Pro as reasoning LLMs without tools to mirror their use in the Copilot Agent harness in \cref{subsection:compare_agent_llm}.
Due to cost constraints, we only run these models on $15$ programs (five math/string/array); in total these experiments cost \$952 USD.
When controlling for model (GPT and Gemini), the trend remains consistent across most axes: LLM accuracy degrades more with larger $N$ compared to agents (Figure~\ref{fig:gemini_gpt}).
This further backs our claim that tool use is what allows agents to excel in \model{} tasks, not differences in underlying model abilities.

\subsection{Further Tool Call Analysis} \label{appendix:tool_breakdown}

We now provide more analyses on agent tool calls.
First, we report the names of tool calls agents use in Table~\ref{tab:tool-call-summary}.
The three most common tool types are consistently \inlinecode{edit}, \inlinecode{bash}, and \inlinecode{view}, which are also popular tool calls in Codex and Copilot \citep{zhu2026tracelab}.

We also decompose the analysis from Figure~\ref{fig:cabra_trace} across the six models in our agent harness, showing the output tokens each model spends on different tool call types.
Our analysis is mostly consistent with our three main trends: 1) function traversal and function search often elicit more \textsc{Analyze} tokens not \textsc{Edit}; 2) runtime resolution elicits near-constant tokens of all tool call types; and 3) instruction following elicits more \textsc{Edit} tokens---but often times more \textsc{Analyze} tokens as well.

\subsection{SWE-Bench Analysis with Per-Model Breakdown} \label{appendix:swe_bench}

We now provide model-specific analyses for the SWE-Bench tool call breakdowns and correlations in \cref{subsection:swe_bench}. We use the following model endpoints in the format \texttt{harness/model}:
\begin{itemize}
    \item \texttt{mini-v2.0.0/claude-4-5-opus-high}
    \item \texttt{mini-v2.0.0/claude-4-5-sonnet-high}
    \item \texttt{mini-v2.0.0/gpt-5-2-high}
    \item \texttt{mini-v2.0.0/gpt-5-mini}    
    \item \texttt{mini-v2.0.0/gemini-3-flash-high}
    \item \texttt{mini-v1.15.0/gemini-3-pro-preview-20251118}
\end{itemize}

All models use the SWE-Agent Mini harness \texttt{mini-v2.0.0} \citep{yang2024sweagent}, except Gemini-3 Pro, as results on this harness are corrupted (the model fails every task).

\begin{figure}
    \centering
    \includegraphics[width=\linewidth]{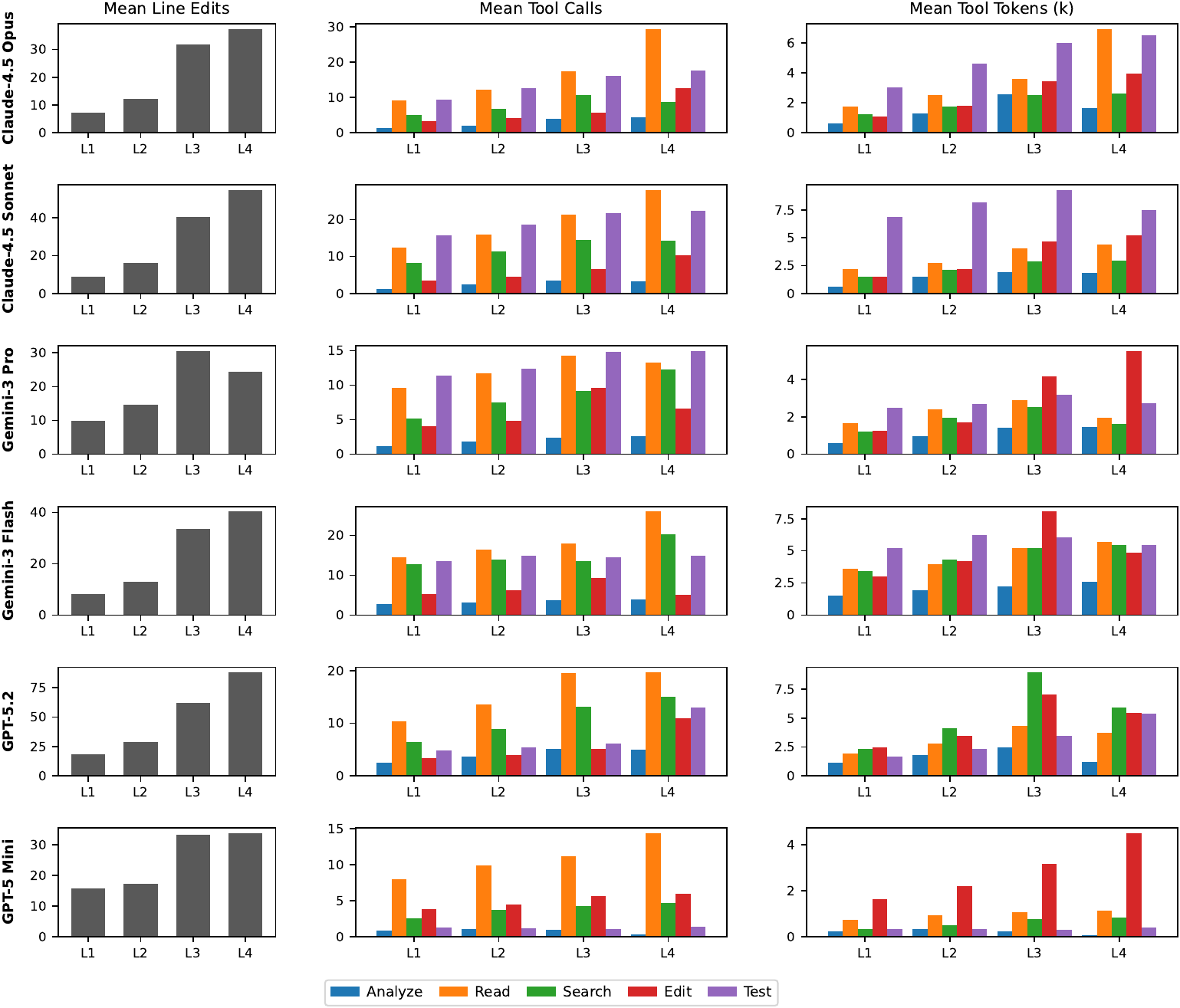}
    \caption{Number of edits, tool calls, and tokens spent on tools calls in Figure~\ref{fig:swe_bench} (left) segmented by agent. As human-annotated difficulty levels rise ($L1 \rightarrow ... \rightarrow L4$), agents do not only edit more code---they also read, analyze, or test more.}
    \label{fig:swe_bench_tool_breakdown}
\end{figure}

We present a per-model breakdown of Figure~\ref{fig:swe_bench} (left) in Figure~\ref{fig:swe_bench_tool_breakdown}, showing that our trend persists across models: they do not only spend more tokens on \textsc{Edit} operations.
To account for model-specific differences in our correlational analysis (Figure~\ref{fig:swe_bench}, right), we fit a separate logistic regression for each metric $m$:

\begin{equation}
\operatorname{logit}\!\left[
\Pr(\mathrm{score}_i = 1 \mid \mathrm{model}_i, m_i)
\right]
= \beta_0 + C(\mathrm{model}_i) + \beta_m m_i,
\label{eq:regression}
\end{equation}

where $score \in {0,1}$ denotes whether the agent completed the task, and $C(model)$ controls for agent baseline success.
For each metric, we compare the augmented regression in Equation~\ref{eq:regression} with a baseline regression using just $C(model)$.
We measure improvement in model fit via the likelihood-ratio (LR) statistic:

\begin{equation}
2\left(\ell_{\mathrm{augmented}}-\ell_{\mathrm{baseline}}\right),
\end{equation}

where $\ell_{\mathrm{augmented}}$ and $\ell_{\mathrm{baseline}}$ are the maximized log-likelihoods of the two models.
Larger LR indicates $m$ improves fit beyond model identity alone.
Edit volume is not the most predictive signal for coding agent accuracy (Table~\ref{table:metric_breakdown})---mirroring the trend in Figure~\ref{fig:swe_bench} (right).
Since \textsc{Search} is also arguably a tool call for understanding, we also run regressions to predict agent accuracy from tool call tokens and count when combining \textsc{Read}, \textsc{Search}, and \textsc{Analyze}.
These metrics are still more predictive of agent accuracy than edit volume.

\subsection{Studying the Impact of Agent Collaboration on our Evaluations} \label{appendix:sandbox}

When reviewing our agent traces, we found cases where our agents attempted to get information from other parallel runs (recalling the OpenAI security incident reported with Hugging Face \citep{openai2026huggingface}).
In particular, we found that some agents used the Copilot CLI session store database to search for similar sessions (although this was very rare; only 72 / $\sim$43k traces). Others explicitly searched the local filesystem for other agents' \texttt{solution.py}. Attempts to read these files were permission-denied; however, all agents shared access to \texttt{/tmp}, which some used for intermediate tasks. One example of an agent looking for other solutions (this tool call was permission-denied):

\lstdefinelanguage{json}{
    basicstyle=\ttfamily\scriptsize,
    numbers=none,
    breaklines=true,
    showstringspaces=false,
    frame=single,
    rulecolor=\color{black!15},
    backgroundcolor=\color{black!2},
    string=[s]{"}{"},
    stringstyle=\color{black!75},
    keywordstyle=\color{black!75},
    commentstyle=\color{black!50},
    columns=fullflexible,
    keepspaces=true,
    aboveskip=0.7em,
    belowskip=0.7em
}

\begin{lstlisting}[language=json]
{
  "command": "find /input_dir/ -name \"solution.py\" -type f -readable 2>/dev/null | head -n 20",
  "description": "Find readable solution.py files"
}
\end{lstlisting}

\begin{figure}
    \centering
    \includegraphics[width=\linewidth]{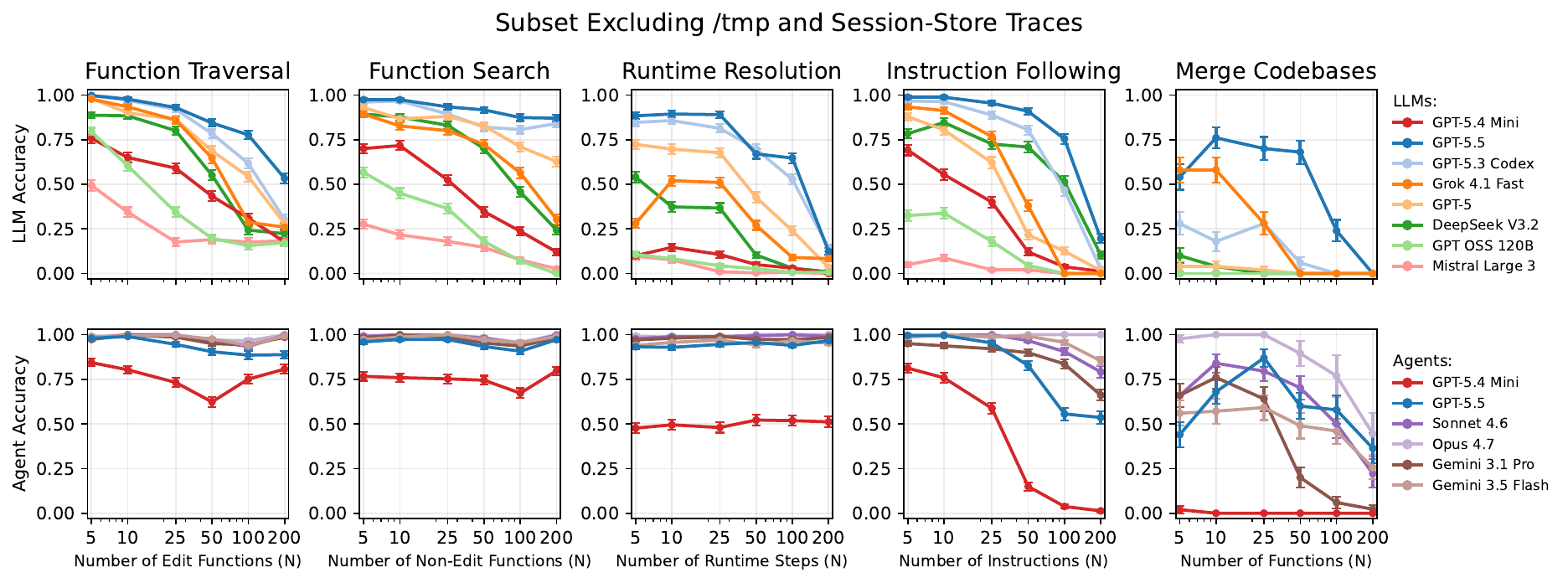}
    \caption{Agent accuracy on \model{} tasks after we filter out suspicious runs where agents may attempt to communicate with other agents. The trend mirrors Figure~\ref{fig:llm_vs_agent} and \Cref{fig:merge_code}b: agent accuracy typically only scales with task size for the merge codebases task that demands more understanding.}
    \label{fig:sandbox_filter_accuracy}
\end{figure}

\begin{figure}
    \centering
    \includegraphics[width=\linewidth]{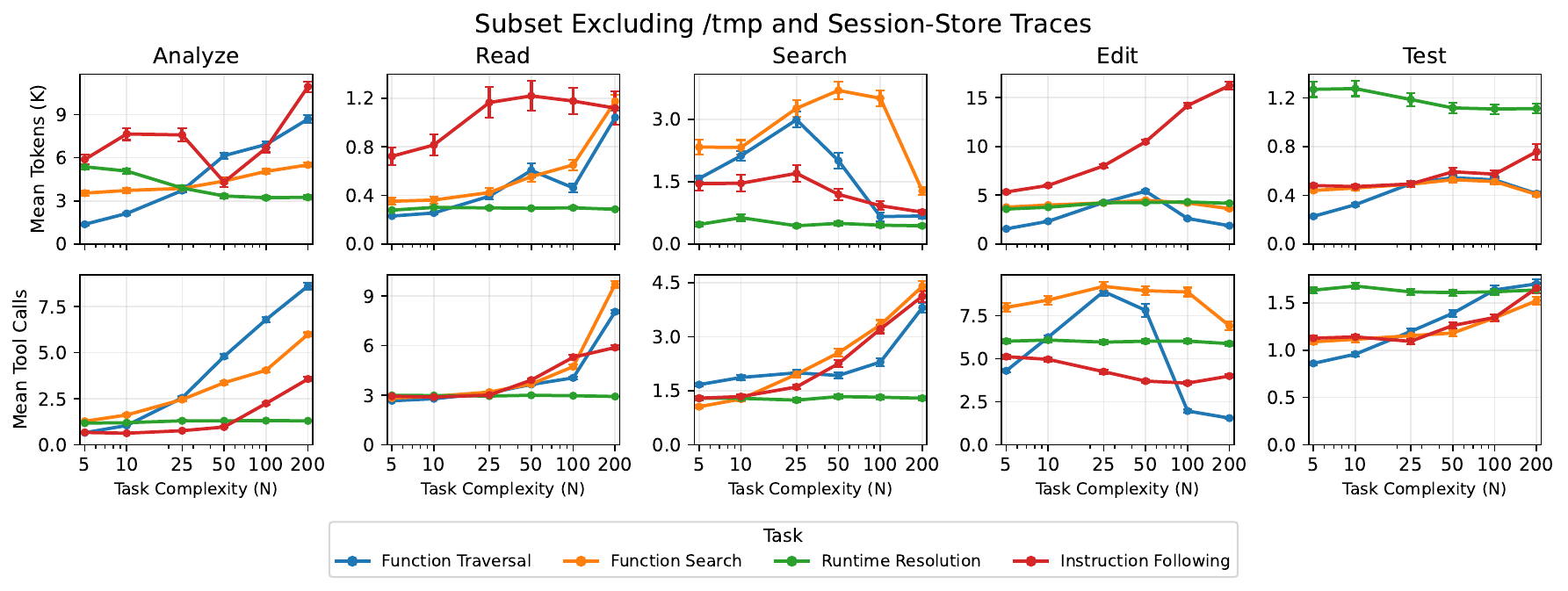}
    \caption{Comparison of tool use on \model{} tasks after we filter out suspicious runs where agents may attempt to communicate with other agents. The trend mirrors Figure~\ref{fig:cabra_trace}: tool use for \textsc{Analyze} and \textsc{Read} more consistently scale than \textsc{Edit} for our initial tasks in \cref{section:static}, and tool calls for \textsc{Analyze} balloon for the merge codebases task in \cref{section:understanding}.}
    \label{fig:sandbox_filter_tools}
\end{figure}

To ensure this does not impact our results, we first examine agent accuracy (Figure~\ref{fig:sandbox_filter_accuracy}) and tool use (Figure~\ref{fig:sandbox_filter_tools}) after filtering traces that could have potentially accessed other runs (i.e., accessed the \texttt{/tmp} folder with other agent solutions or used the \inlinecode{session\_store\_sql} tool).
This filters just 11\% of our traces (mostly innocuous uses of \texttt{/tmp}), and trends are consistent with our main results.

Next, we use a Docker image that exposes only the input codebase to better sandbox the agent (preventing any filesystem-based cross-trace leak). We set up the Docker image to be minimal but provide tools the agent might find useful:

\begin{center}
\begin{lstlisting}[style=bash]
FROM node:22-bookworm-slim

RUN apt-get update \
    && apt-get install --yes --no-install-recommends \
        bash \
        build-essential \
        ca-certificates \
        coreutils \
        curl \
        diffutils \
        file \
        findutils \
        gawk \
        grep \
        jq \
        less \
        patch \
        procps \
        python3 \
        python3-pip \
        python3-venv \
        ripgrep \
        sed \
        tree \
        wget \
    && npm install --global "@github/copilot@1.0.81" \
    && apt-get clean \
    && rm -rf /var/lib/apt/lists/* \
    && useradd --create-home --uid 10001 agent

USER 10001:10001
WORKDIR /workspace

RUN copilot --version
\end{lstlisting}
\end{center}

Using this file, we script Copilot \texttt{v1.0.81} (run in September 2026) to operate inside the Docker image while also disabling the \inlinecode{session\_store\_sql} tool to prevent unecessary searches:

\begin{center}
\begin{lstlisting}[style=bash]
docker create \
  --name <container_name> \
  --label cabra.agent=copilot \
  --network bridge \
  --init \
  --mount type=bind,source=<workdir>,target=/workspace \
  --workdir /workspace \
  --user 10001:10001 \
  --cap-drop ALL \
  --security-opt no-new-privileges \
  --memory 32g \
  --env HOME=/home/agent \
  --env COPILOT_GITHUB_TOKEN \
  <docker_image> \
  copilot \
  --model <model_name> \
  -C /workspace \
  -p <prompt> \
  --mode autopilot \
  --silent \
  --allow-all-tools \
  --excluded-tools=session_store_sql \
  --deny-tool=session_store_sql \
  --session-id=<session_id> \
  --no-ask-user \
  --no-remote \
  --no-remote-export
\end{lstlisting}
\end{center}

\begin{figure}
    \centering
    \includegraphics[width=\linewidth]{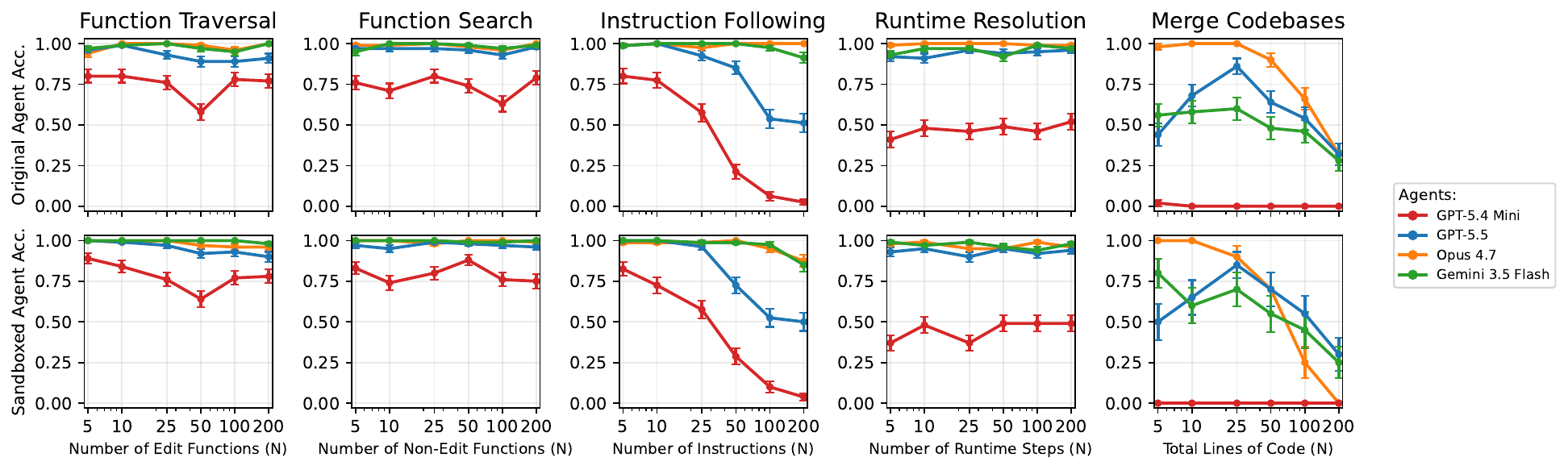}
    \caption{Comparison of agent accuracy on initial \model{} tasks in a fully sandboxed environment versus our reported non-sandboxed runs. Trends remain consistent: agent accuracy often does not degrade as task size increases---except for our merge codebases task which targets understanding.}
    \label{fig:sandbox_accuracy}
\end{figure}

\begin{figure}
    \centering
    \includegraphics[width=\linewidth]{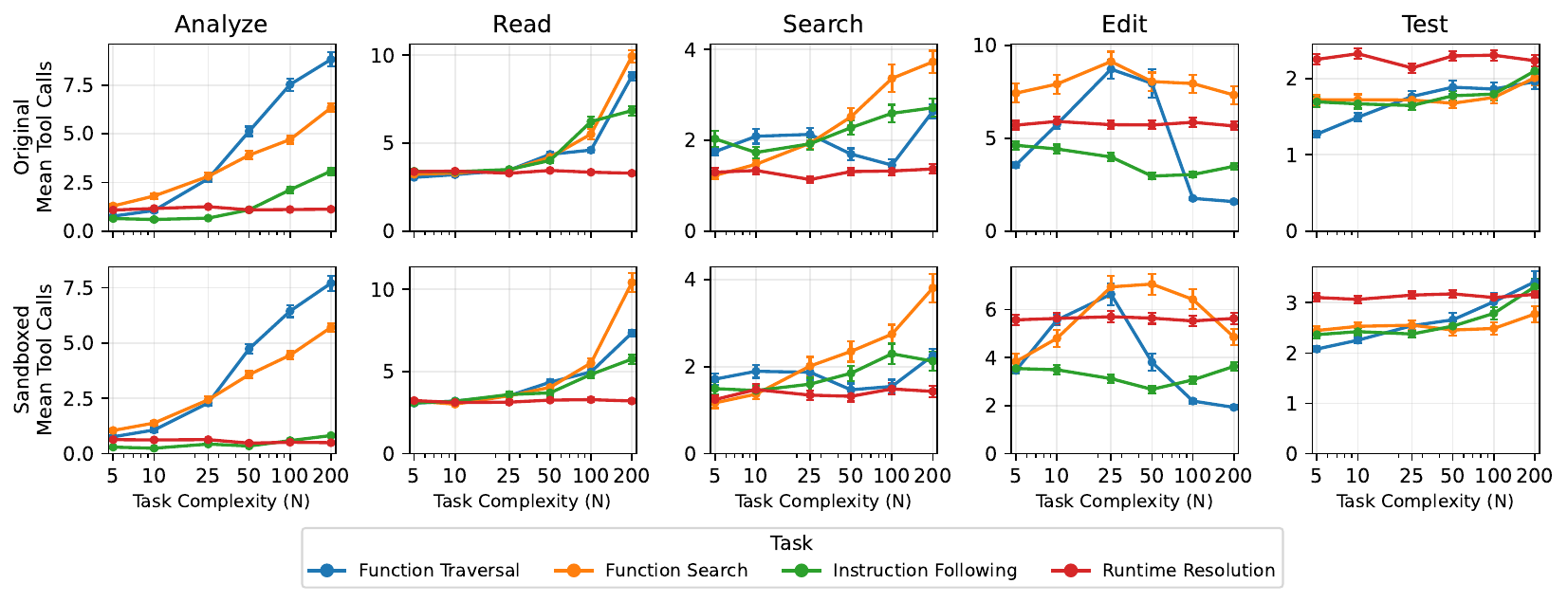}
    \caption{Comparison of tool use on initial \model{} tasks in a fully sandboxed environment versus our reported non-sandboxed runs. Trends remain consistent: tool use for \textsc{Edit} do not always scale, but \textsc{Read} and \textsc{Analyze} tools do more consistently.}
    \label{fig:sandbox_tool}
\end{figure}

Using this approach, we run $n=20$ examples using the math version of our codebases across all tasks, with Opus-4.7, GPT-5.5, GPT-5.4 Mini, and Gemini-3.5 Flash (cost breakdown in \Cref{tab:sandboxed-cost}).
We present accuracy in Figure~\ref{fig:sandbox_accuracy} and tool use metrics in Figure~\ref{fig:sandbox_tool} which are also consistent with our main results: agent accuracy mostly degrades with task size for our understanding-intensive merge codebases task, and usage of \textsc{Read} and \textsc{Analyze} tools scale beyond just \textsc{Edit} with task size---with \textsc{Analyze} growing especially in the merge codebases task.

Overall, as (1) filtering traces with potential cross-agent communication; and (2) re-running agents with problematic tools disabled in contained Docker images both did not impact our conclusions, we are  confident that the main trends we present in this paper are unaffected by cross-agent communication.

\subsection{Editing Tasks Without Much Understanding} \label{appendix:edit_without_understanding}

Our evaluations with \model{} in \cref{section:results} reveal two task types where scaling complexity changes behavior in agent tool calls: 1) \textsc{Understand} tool calls increase while \textsc{Edit} tool calls remain stable (i.e., function search and function traversal); and 2) Both \textsc{Understand} and \textsc{Edit} tool calls increase (i.e., instruction following).
To confirm that understanding and editing are separate constructs, we now design a task that scales editing requirements with complexity without large changes in understanding requirements. 

We study two different tasks:
\begin{enumerate}
    \item A comment copying task which uses the \model{} tasks in \cref{section:static} similar to instruction-following, but the model is asked to add a comment with $n$ words (Prompt~\ref{prompt:instructions_edit_only})
    \item A variable renaming task where a model receives a simple calculation with $200$ operations in a block of code (Code~\ref{code:rename_variables}) and must rename $n$ variables as defined in the prompt (Prompt~\ref{prompt:replace_vars}).
\end{enumerate}

\begin{figure}
    \centering
    \includegraphics[width=\linewidth]{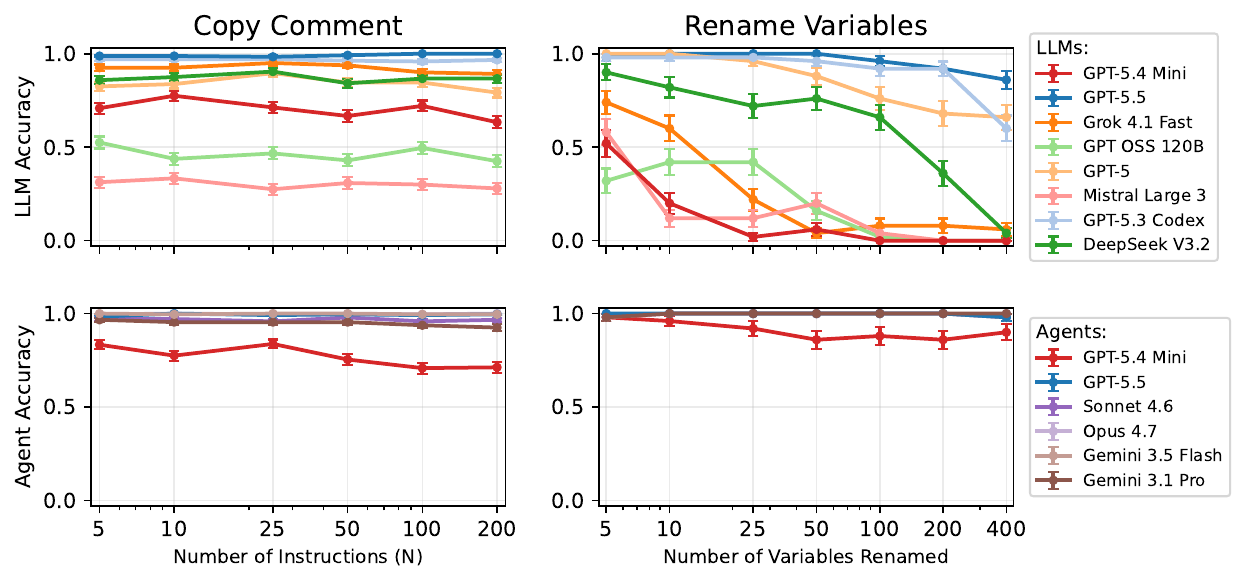}
    \caption{Evaluation of our synthetic tasks that: 1) demand adding comments of increased words $n$; and 2) renaming $n$ variables in a block of code. As $n$ increases, LLM and agent accuracy remains stable for (1), but in (2), LLM accuracy decreases while agent accuracy remains stable.}
    \label{fig:edit_no_understand_accuracy}
\end{figure}

\begin{figure}
    \centering
    \includegraphics[width=\linewidth]{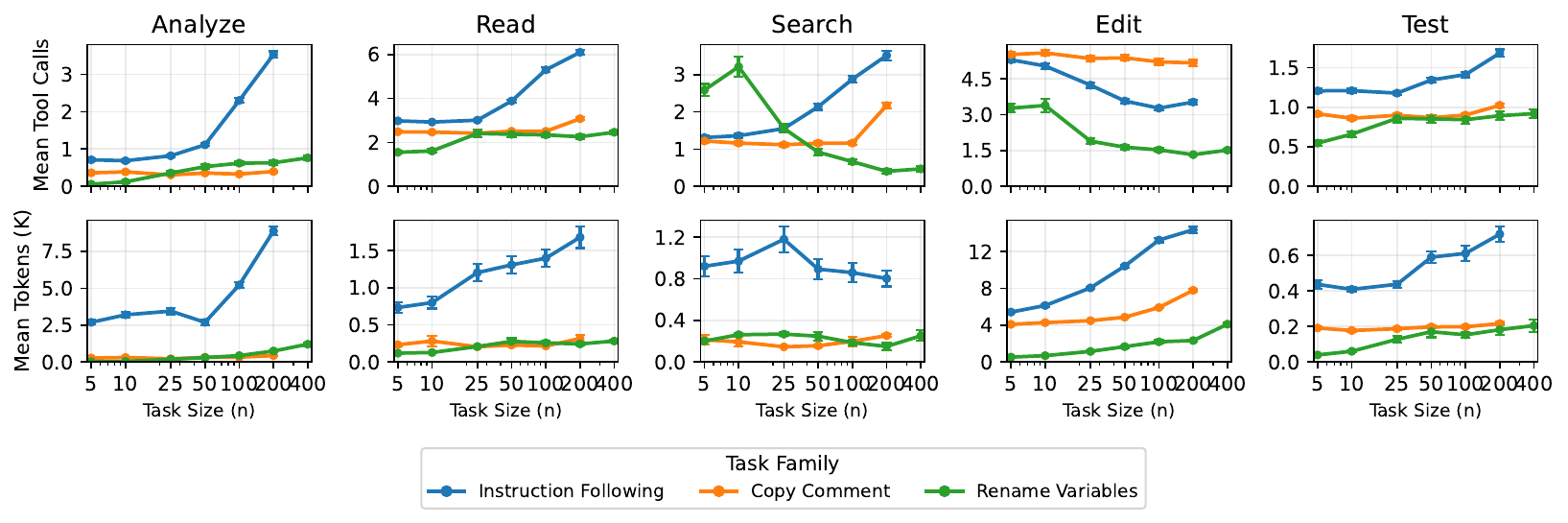}
    \caption{Tool call categorization of the instruction-following task in Figure~\ref{fig:cabra_trace} versus tasks where models: 1) copy comments with $n$ words; and 2) rename $n$ variables in a block of code. As $n$ increases, output tokens for \textsc{Edit} scales for (1) and (2), but \textsc{Analyze} and \textsc{Read} scale modestly compared to instruction-following. Thus, there exist tasks that can increase editing requirements without largely increasing understanding requirements.}
    \label{fig:edit_no_understand_tool_calls}
\end{figure}

Evaluations on this task reveal that as $n$ increases: 1) LLM and agent accuracy remains consistent for comment copying, but LLM accuracy drops in variable renaming with $n$ (Figure~\ref{fig:edit_no_understand_accuracy}); and 2) \textsc{Edit} tool calls scale in both tasks without large changes in \textsc{Understand} tool calls, especially compared to our first instruction-following task in \model{} (Figure~\ref{fig:edit_no_understand_tool_calls}).
The number of \textsc{Edit} tool calls does not scale for any task, but one can arbitrarily make a require more \textsc{Edit} tool calls by forcing models to make smaller edits at a time.

Overall, this further backs our argument that understanding and editing are distinct coding agent abilities that can be uniquely targeted and measured in evaluations.

\subsection{Additional Results on Extracting Equivalent Code} \label{appendix:merge}

We first show both LLM and coding agent accuracy on the merge codebases task (\cref{section:understanding}).
Both model types degrade in accuracy with large $n$, but many LLMs fail to ever clear $0.5$ accuracy---confirming the task's difficulty. 

\begin{figure}
    \centering
    \includegraphics[width=\linewidth]{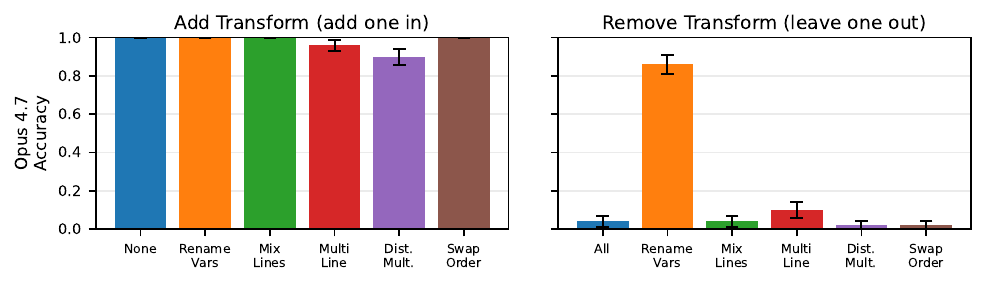}
    \caption{Evaluation of transformation type ablations in Figure~\ref{fig:ablation} (right) for $n=200$ with Opus 4.7---our strongest agent. The trend follows Figure~\ref{fig:merge_code_ablate_transform_model}: Opus struggles more with the multi-line and distributive transformations and sees the largest accuracy improvements when variables are not renamed.
    }
    \label{fig:merge_code_ablate_opus}
\end{figure}

\begin{figure}
    \centering
    \includegraphics[width=\linewidth]{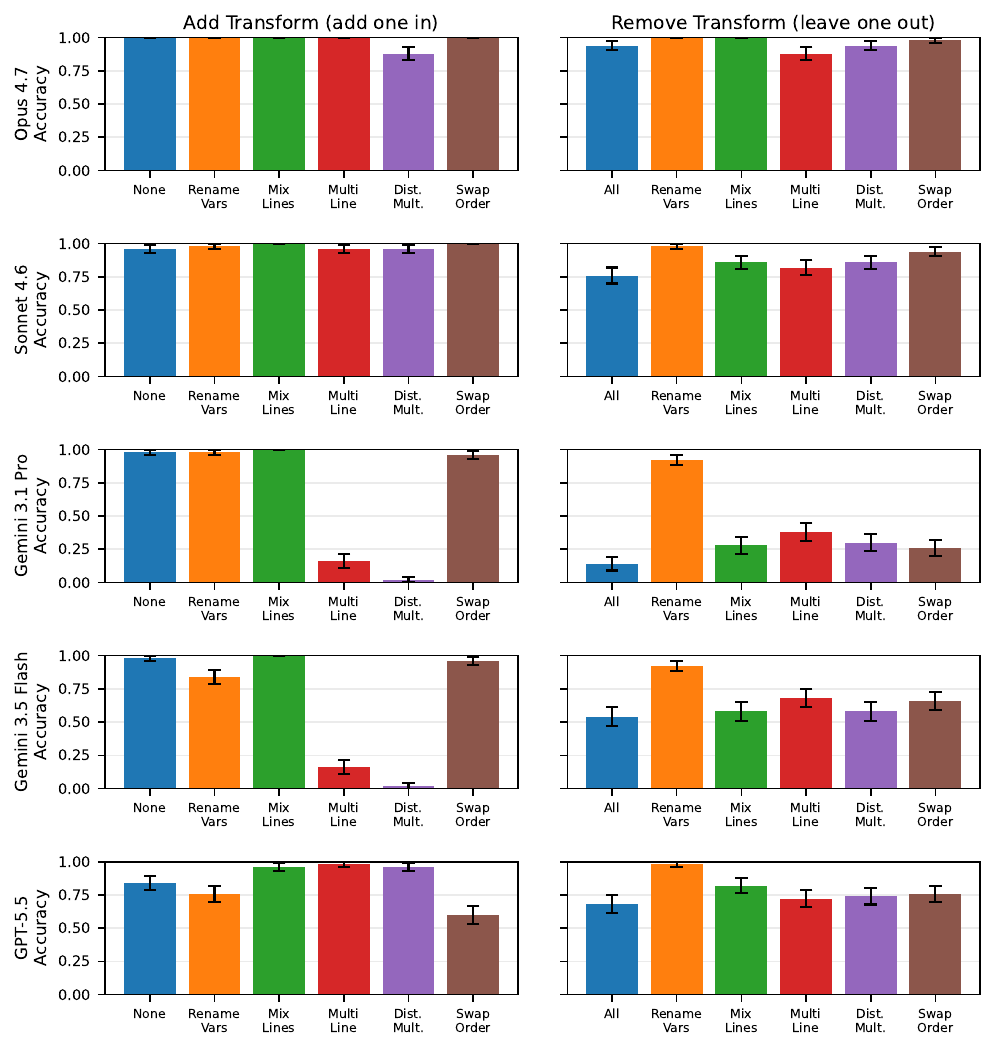}
    \caption{Evaluation of transformation type ablations in Figure~\ref{fig:ablation} (right) segmented by agent $(n=50)$. Gemini and Opus 4.7 struggle more with the multi-line and distributive transformations, while GPT-5 struggles with the swap order transformation. All models see the largest accuracy improvements when comments are not renamed, further supporting that models may over-rely on surface-level naming over behavior when understanding code.
    }\label{fig:merge_code_ablate_transform_model}
\end{figure}

We also present the follow-up analyses in Figure~\ref{fig:ablation} segmented across our six agents.
The add-one-in ablation accuracy results are model-specific, but 3/6 models struggle with ones related to multi-line differences (Figure~\ref{fig:merge_code_ablate_transform_model}, left).
In the leave-one-out ablations, results are consistent: all models improve most when variables are not renamed (Figure~\ref{fig:merge_code_ablate_transform_model}, right), which could suggest over-reliance on surface-level naming when inferring code behavior. 
Since Opus-4.7 saturates the task at $n=50$, we repeat this experiment for $n=200$ with Opus alone in Figure~\ref{fig:merge_code_ablate_opus}, which shows similar trends: the model struggles with multi-line differences and variable renaming.

\begin{figure}
    \centering
    \includegraphics[width=\linewidth]{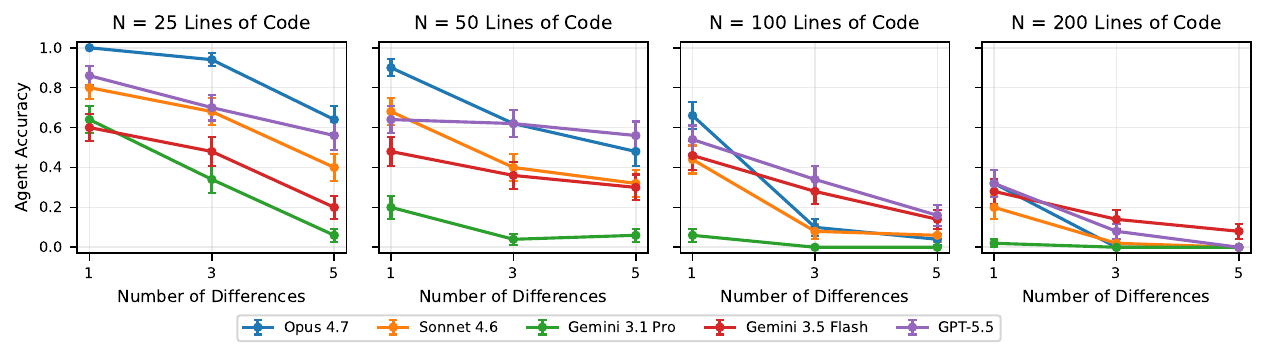}
    \caption{Evaluation of increased number of differences in Figure~\ref{fig:ablation} (right) segmented by agent. The trend is consistent: agents struggle to identify increased number of differences.
    }
    \label{fig:merge_code_ablate_diff_model}
\end{figure}

\begin{figure}
    \centering
    \includegraphics[width=\linewidth]{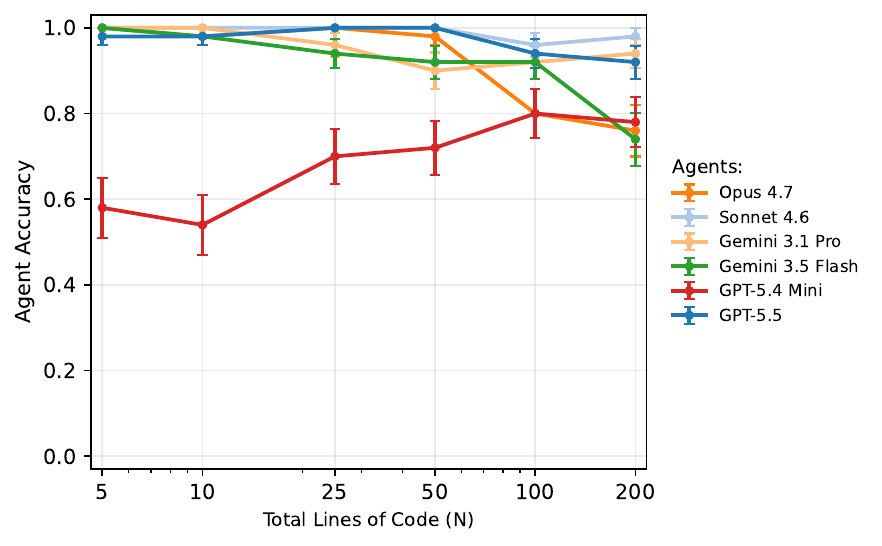}
    \caption{Evaluation of the extract logic task in \cref{section:understanding} but instructing agents that the optimal solution is exactly a one-line difference. Agent accuracy is much higher compared to Figure~\ref{fig:merge_code}b, showing difficulty lies in navigating the search space to find the optimal solution with minimal edits.
    }
    \label{fig:merge_code_solution}
\end{figure}

Finally, the trend for locating increasing number of different lines is consistent across models: all agents struggle more when they must identify more one-line differences (Figure~\ref{fig:merge_code_ablate_diff_model}).
This also suggests that difficulty arises from navigating the search space of all possible solutions to find the most efficient one (i.e., least lines of code in child classes).
To validate this, we run the same experiment as Figure~\ref{fig:merge_code}b but instruct models that there is only a single one-line difference; as expected, agent accuracy is much higher (Figure~\ref{fig:merge_code_solution}).

\subsection{Prompts} \label{appendix:prompts}

We provide prompts with our problem statements for Remove Dead Code (Prompt~\ref{prompt:remove_dead_code}), Add Parameter (Prompt~\ref{prompt:add_parameter}), Add Return Value (Prompt~\ref{prompt:add_return}), Cache Function (Prompt~\ref{prompt:cache_function}), Don't Repeat Yourself (Prompt~\ref{prompt:dont_repeat}), and Merge Codebases (Prompt~\ref{prompt:merge_code}) tasks.
We also show how we minimally tailor these prompts for the string (Prompt~\ref{prompt:add_parameter_string}) and array (Prompt~\ref{prompt:add_parameter_array}) versions of our code.
Prompt~\ref{prompt:runtime} has the prompt for tasks involving scaling runtime complexity, and Prompts~\ref{prompt:instructions} and \ref{prompt:instructions_edit_only} for instruction following tasks with and without scaling understanding tool calls, respectively.

For \cref{subsection:agent}, Prompt~\ref{prompt:llm_full} and \ref{prompt:llm_diff} are used for evaluating LLMs when they generate the full codebase or diff patches, respectively.
Prompts~\ref{prompt:classify_cabra} and \ref{prompt:classify_swebench} are the classification prompts of model tool calls on our synthetic \model{} tasks and SWE-Bench, respectively.

\clearpage

\begin{algorithm}[t]
\caption{Sampling Procedure for Directed Acyclic Call Graphs (\cref{subsection:dag})}
\label{alg:sample-dag}
\begin{algorithmic}[1]
\Require Number of vertices $N$
\Ensure A directed acyclic graph $G=(V,E)$

\vspace{1em}
\Statex \textbf{Step 1: Sample the number of topological levels} \vspace{1em}

\State Sample $h \sim \mathrm{Uniform}(0.45,0.55)$
\State $L \gets \max\{1,\operatorname{round}(N^h)\}$

\vspace{1em}
\Statex \textbf{Step 2: Distribute vertices across levels} 
\vspace{1em}

\State $V_{\mathrm{left}} \gets N$
\State $V \gets \varnothing$, $E \gets \varnothing$

\For{$\ell \gets 1$ to $L$}
    \State $L_{\mathrm{left}} \gets L-\ell+1$
    \State $\lambda_\ell \gets V_{\mathrm{left}}/L_{\mathrm{left}}$
        \Comment{Poisson mean based on nodes left}

    \If{$\ell=L$}
        \State $n_\ell \gets V_{\mathrm{left}}$
            \Comment{Assign all remaining vertices}
    \Else
        \State $\widetilde n_\ell \sim \mathrm{Poisson}(\lambda_\ell)$
        \State $n_\ell \gets
        \min\!\left\{
            \max\{1,\widetilde n_\ell\},
            V_{\mathrm{left}}-(L_{\mathrm{left}}-1)
        \right\}$
        \Comment{At least one vertex per level}
    \EndIf

    \State Create level $V_\ell$ containing $n_\ell$ new vertices
    \State $V \gets V \cup V_\ell$
    \State $V_{\mathrm{left}} \gets V_{\mathrm{left}}-n_\ell$
\EndFor

\vspace{1em}
\Statex \textbf{Step 3: Sample one incoming edge per non-root vertex} 
\vspace{1em}

\For{$\ell \gets 2$ to $L$}
    \ForAll{$v \in V_\ell$}
        \State Sample $\widetilde\Delta \sim \mathrm{Geom}(0.9)$
            \Comment{Prefer nearby levels}
        \State $\Delta \gets \min\{\widetilde\Delta,\ell-1\}$
            \Comment{Clip to an existing earlier level}
        \State Sample $u \sim \mathrm{Uniform}(V_{\ell-\Delta})$
            \Comment{Choose a parent within that level}
        \State $E \gets E \cup \{(u,v)\}$
    \EndFor
\EndFor

\State \Return $G=(V,E)$
\end{algorithmic}
\end{algorithm}

\begin{table*}[t]
\centering
\caption{Input and output tokens (in millions) used by our final evaluation runs by model and task family. We calculate costs in USD from LLM API providers. Cached tokens are included in the ``In'' column but accounted for separately in the USD cost. Total cost: \$28,262.}
\label{table:cost}
\scriptsize
\setlength{\tabcolsep}{3pt}
\resizebox{\textwidth}{!}{%
\begin{tabular}{lrrrrrrrrrrrrr}
\toprule
Model & \multicolumn{2}{c}{Traversal} & \multicolumn{2}{c}{Instruction} & \multicolumn{2}{c}{Search} & \multicolumn{2}{c}{Runtime} & \multicolumn{2}{c}{Merge} & \multicolumn{3}{c}{Overall} \\
\cmidrule(lr){2-3} \cmidrule(lr){4-5} \cmidrule(lr){6-7} \cmidrule(lr){8-9} \cmidrule(lr){10-11} \cmidrule(lr){12-14}
 & In & Out & In & Out & In & Out & In & Out & In & Out & In & Out & Cost (USD) \\
\midrule
\multicolumn{14}{l}{\textit{LLMs}} \\
DeepSeek-V3.2 & 7.0 & 17.4 & 5.9 & 22.1 & 7.1 & 18.8 & 5.1 & 22.2 & 1.0 & 3.6 & 26.2 & 84.1 & \$41 \\
Mistral-Large-3 & 6.9 & 7.6 & 6.0 & 14.6 & 6.9 & 8.0 & 5.1 & 5.3 & 1.0 & 1.7 & 26.0 & 37.2 & \$69 \\
gpt-5.3-codex & 6.6 & 9.3 & 5.8 & 11.8 & 6.7 & 9.2 & 5.0 & 12.4 & 1.0 & 1.7 & 25.0 & 44.4 & \$665 \\
gpt-5.4-mini & 6.6 & 19.0 & 5.8 & 24.1 & 6.7 & 16.4 & 5.0 & 18.9 & 1.0 & 2.6 & 25.0 & 80.9 & \$383 \\
gpt-5.5 & 6.6 & 8.5 & 5.8 & 13.6 & 6.7 & 8.6 & 5.0 & 11.3 & 1.0 & 1.8 & 25.0 & 43.9 & \$1,440 \\
gpt-5 & 6.6 & 15.1 & 5.8 & 19.8 & 6.7 & 15.8 & 5.0 & 21.7 & 1.0 & 1.8 & 25.0 & 74.2 & \$773 \\
gpt-oss-120b & 6.7 & 11.9 & 5.9 & 14.0 & 6.8 & 14.1 & 5.1 & 13.4 & 1.0 & 1.2 & 25.5 & 54.6 & \$10 \\
grok-4-1-fast-reasoning & 6.6 & 2.8 & 5.7 & 4.3 & 6.6 & 4.4 & 5.0 & 2.9 & 1.0 & 0.4 & 25.0 & 14.9 & \$12 \\
\midrule
\multicolumn{14}{l}{\textit{Agents}} \\
claude-opus-4.7 & 1874.4 & 39.1 & 1229.6 & 48.9 & 2527.1 & 43.6 & 1452.0 & 38.9 & 468.1 & 15.6 & 7551.1 & 186.0 & \$10,407 \\
claude-sonnet-4.6 & 1309.3 & 39.1 & 1078.5 & 57.7 & 2055.5 & 38.5 & 1108.4 & 35.2 & 308.2 & 20.5 & 5859.9 & 191.1 & \$5,822 \\
gemini-3.1-pro-preview & 711.8 & 5.4 & 710.8 & 12.8 & 628.5 & 4.6 & 450.7 & 5.8 & 598.4 & 5.6 & 3100.2 & 34.2 & \$1,950 \\
gemini-3.5-flash & 1701.9 & 9.3 & 1675.6 & 19.7 & 1552.7 & 9.3 & 1141.2 & 9.0 & 1033.8 & 6.1 & 7105.1 & 53.4 & \$2,775 \\
gpt-5.4-mini & 849.2 & 23.2 & 666.3 & 27.7 & 838.6 & 20.4 & 469.9 & 17.0 & 53.1 & 2.5 & 2877.0 & 90.7 & \$873 \\
gpt-5.5 & 359.9 & 8.4 & 304.7 & 13.3 & 376.7 & 8.7 & 297.4 & 7.3 & 83.1 & 2.8 & 1421.7 & 40.5 & \$3,042 \\
\bottomrule
\end{tabular}%
}
\end{table*}

\begin{table}[h]
    \centering
    \caption{Sandboxed runs cost breakdown using public pricing data (as of September 2026). ``In'' includes cached input, but those tokens are accounted for separately in cost estimates. Total cost: \$3,624.}
    \label{tab:sandboxed-cost}
    \scriptsize
\resizebox{\textwidth}{!}{%
\begin{tabular}{lrrrrrrrrrrrrr}

\toprule
Model & \multicolumn{2}{c}{Traversal} & \multicolumn{2}{c}{Instruction} & \multicolumn{2}{c}{Search} & \multicolumn{2}{c}{Runtime} & \multicolumn{2}{c}{Merge} & \multicolumn{3}{c}{Overall} \\
\cmidrule(lr){2-3} \cmidrule(lr){4-5} \cmidrule(lr){6-7} \cmidrule(lr){8-9} \cmidrule(lr){10-11} \cmidrule(lr){12-14}
 & In & Out & In & Out & In & Out & In & Out & In & Out & In & Out & Cost (USD) \\
\midrule
\multicolumn{14}{l}{\textit{Agents}} \\
claude-opus-4.7 & 318.9 & 4.6 & 129.2 & 5.8 & 415.8 & 4.8 & 342.1 & 4.7 & 82.8 & 2.5 & 1288.9 & 22.4 & \$1,532 \\
gemini-3.5-flash & 630.3 & 3.0 & 517.3 & 6.3 & 569.6 & 3.1 & 446.0 & 2.7 & 318.2 & 2.6 & 2481.5 & 17.7 & \$896 \\
gpt-5.4-mini & 335.6 & 7.9 & 264.4 & 9.8 & 332.6 & 6.8 & 198.9 & 5.7 & 27.2 & 0.8 & 1158.6 & 31.0 & \$299 \\
gpt-5.5 & 142.6 & 2.9 & 119.1 & 4.7 & 153.5 & 2.9 & 125.3 & 2.5 & 44.3 & 1.2 & 584.7 & 14.2 & \$924 \\
\bottomrule
\end{tabular}}
\end{table}

\begin{figure}
    \centering
    \includegraphics[width=\linewidth]{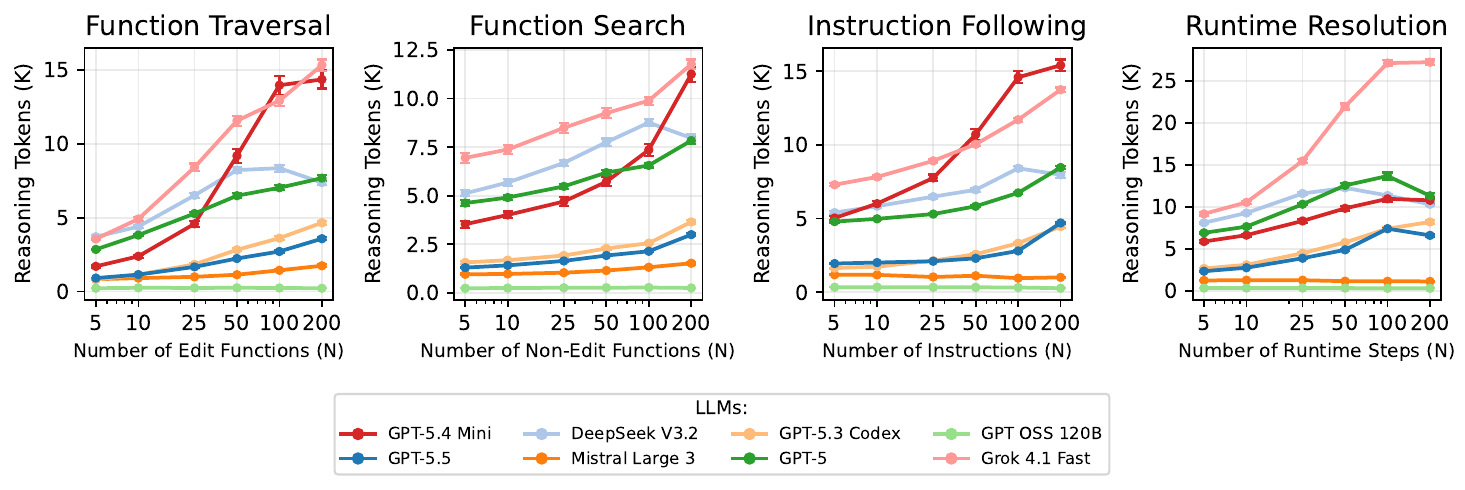}
    \caption{Number of reasoning tokens LLMs use for our four axes of \model{} in \cref{subsection:complexity}: the sum of tokens in \texttt{<thought></thought>} blocks via tiktoken plus the number of reasoning tokens reported by the model provider (in thousands). Reasoning token count typically increases with $N$.}
    \label{fig:llm_reasoning_tokens}
\end{figure}

\begin{figure}
    \centering
    \includegraphics[width=\linewidth]{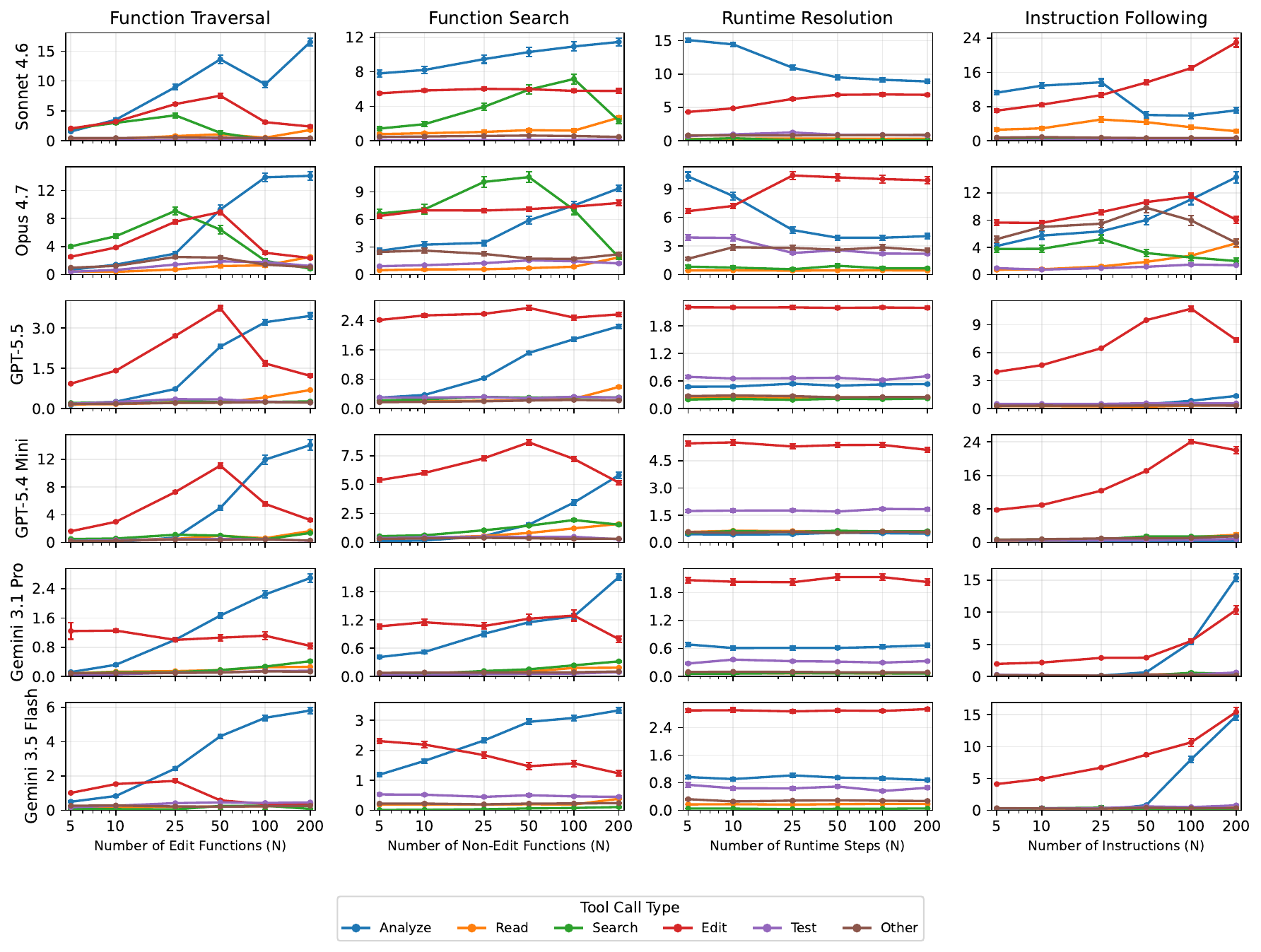}
    \caption{Analysis of model-specific \textbf{output tokens (in thousands) used for tool calls}. Three trends emerge: most models spend more tokens on \textsc{Analyze} in function traversal and search, no tool call type scales in runtime resolution, and instruction-following always requires more \textsc{Edit} tokens (and often more \textsc{Analyze} tokens).}
    \label{fig:tool_call_breakdown_tokens}
\end{figure}

\begin{table}[t]
\centering
\scriptsize
\caption{The ten most frequent tool call names agents call across each complexity axis in \cref{subsection:complexity}. We report the raw number (in thousands) and the proportion of all tools. The model also called \texttt{web\_search} in 48 out of 700k+ cases, which suggests cheating \citep[e.g., the model Googling our prompt verbatim]{yang2026programbench}, but this is rare enough to not impact results.}
\begin{tabular}{llll}
\toprule
Function Traversal & Function Search & Instruction Following & Runtime Resolution \\
\midrule
\texttt{bash} (75.7k; 37.7\%) & \texttt{edit} (82.6k; 35.1\%) & \texttt{bash} (43.8k; 33.8\%) & \texttt{edit} (58.7k; 35.9\%) \\
\texttt{edit} (45.9k; 22.9\%) & \texttt{bash} (62.9k; 26.7\%) & \texttt{view} (24.5k; 18.9\%) & \texttt{bash} (44.4k; 27.2\%) \\
\texttt{view} (31.7k; 15.8\%) & \texttt{view} (36.9k; 15.6\%) & \texttt{edit} (21.9k; 16.9\%) & \texttt{view} (22.5k; 13.8\%) \\
\texttt{complete} (10.8k; 5.4\%) & \texttt{rg} (10.9k; 4.6\%) & \texttt{rg} (8.8k; 6.8\%) & \texttt{complete} (10.8k; 6.6\%) \\
\texttt{create} (8.9k; 4.4\%) & \texttt{complete} (10.8k; 4.6\%) & \texttt{complete} (8.6k; 6.7\%) & \texttt{apply\_patch} (7.2k; 4.4\%) \\
\texttt{rg} (8.4k; 4.2\%) & \texttt{apply\_patch} (9.7k; 4.1\%) & \texttt{glob} (6.6k; 5.1\%) & \texttt{glob} (6.5k; 4.0\%) \\
\texttt{glob} (7.3k; 3.6\%) & \texttt{create} (8.0k; 3.4\%) & \texttt{apply\_patch} (5.7k; 4.4\%) & \texttt{rg} (5.1k; 3.1\%) \\
\texttt{apply\_patch} (5.9k; 2.9\%) & \texttt{glob} (7.3k; 3.1\%) & \texttt{create} (3.6k; 2.7\%) & \texttt{sql} (3.6k; 2.2\%) \\
\texttt{grep} (4.2k; 2.1\%) & \texttt{grep} (4.6k; 2.0\%) & \texttt{sql} (3.4k; 2.6\%) & \texttt{grep} (2.8k; 1.7\%) \\
\texttt{sql} (2.0k; 1.0\%) & \texttt{sql} (1.6k; 0.7\%) & \texttt{grep} (1.6k; 1.2\%) & \texttt{create} (1.9k; 1.2\%) \\
\bottomrule
\end{tabular}
\label{tab:tool-call-summary}
\end{table}

\begin{table}[t]
\centering
\caption{Logistic regression to predict agent's task completion using each tool call metric in \cref{subsection:swe_bench}, controlling for agent. Edit volume is not the most predictive signal for task accuracy.}
\label{table:metric_breakdown}
\begin{tabular}{lrrrr}
\toprule
Metric & Coefficient & Odds Ratio & $p$-value & LR Statistic ($\downarrow$) \\
\midrule
All Tokens & -0.788 & 0.455 & $<0.001$ & 224.31 \\
\textsc{Understand+Search}$_{\mathrm{tok}}$ & -0.659 & 0.517 & $<0.001$ & 189.52 \\
\textsc{Understand}$_{\mathrm{num}}$ & -0.589 & 0.555 & $<0.001$ & 183.63 \\
All Tools & -0.629 & 0.533 & $<0.001$ & 171.46 \\
\textsc{Understand}$_{\mathrm{tok}}$ & -0.564 & 0.569 & $<0.001$ & 156.47 \\
\textsc{Understand+Search}$_{\mathrm{num}}$ & -0.556 & 0.574 & $<0.001$ & 154.33 \\
\textsc{Read}$_{\mathrm{num}}$ & -0.492 & 0.611 & $<0.001$ & 133.21 \\
\textsc{Edit}$_{\mathrm{tok}}$ & -0.468 & 0.626 & $<0.001$ & 115.01 \\
\textsc{Analyze}$_{\mathrm{num}}$ & -0.430 & 0.650 & $<0.001$ & 104.47 \\
\textsc{Search}$_{\mathrm{tok}}$ & -0.449 & 0.638 & $<0.001$ & 95.57 \\
\textsc{Read}$_{\mathrm{tok}}$ & -0.437 & 0.646 & $<0.001$ & 94.45 \\
\textsc{Analyze}$_{\mathrm{tok}}$ & -0.403 & 0.669 & $<0.001$ & 94.14 \\
Lines Edited & -0.457 & 0.633 & $<0.001$ & 85.10 \\
\textsc{Test}$_{\mathrm{tok}}$ & -0.492 & 0.611 & $<0.001$ & 78.68 \\
\textsc{Edit}$_{\mathrm{num}}$ & -0.361 & 0.697 & $<0.001$ & 77.11 \\
\textsc{Test}$_{\mathrm{num}}$ & -0.497 & 0.608 & $<0.001$ & 67.62 \\
\textsc{Search}$_{\mathrm{num}}$ & -0.364 & 0.695 & $<0.001$ & 65.54 \\
\textsc{Other}$_{\mathrm{tok}}$ & -0.170 & 0.843 & $<0.001$ & 15.88 \\
Files Edited & -0.154 & 0.857 & $<0.001$ & 14.89 \\
\textsc{Other}$_{\mathrm{num}}$ & -0.106 & 0.899 & 0.016 & 5.67 \\
\bottomrule
\end{tabular}

\end{table}



\begin{figure}
    \centering
    \includegraphics[width=\linewidth]{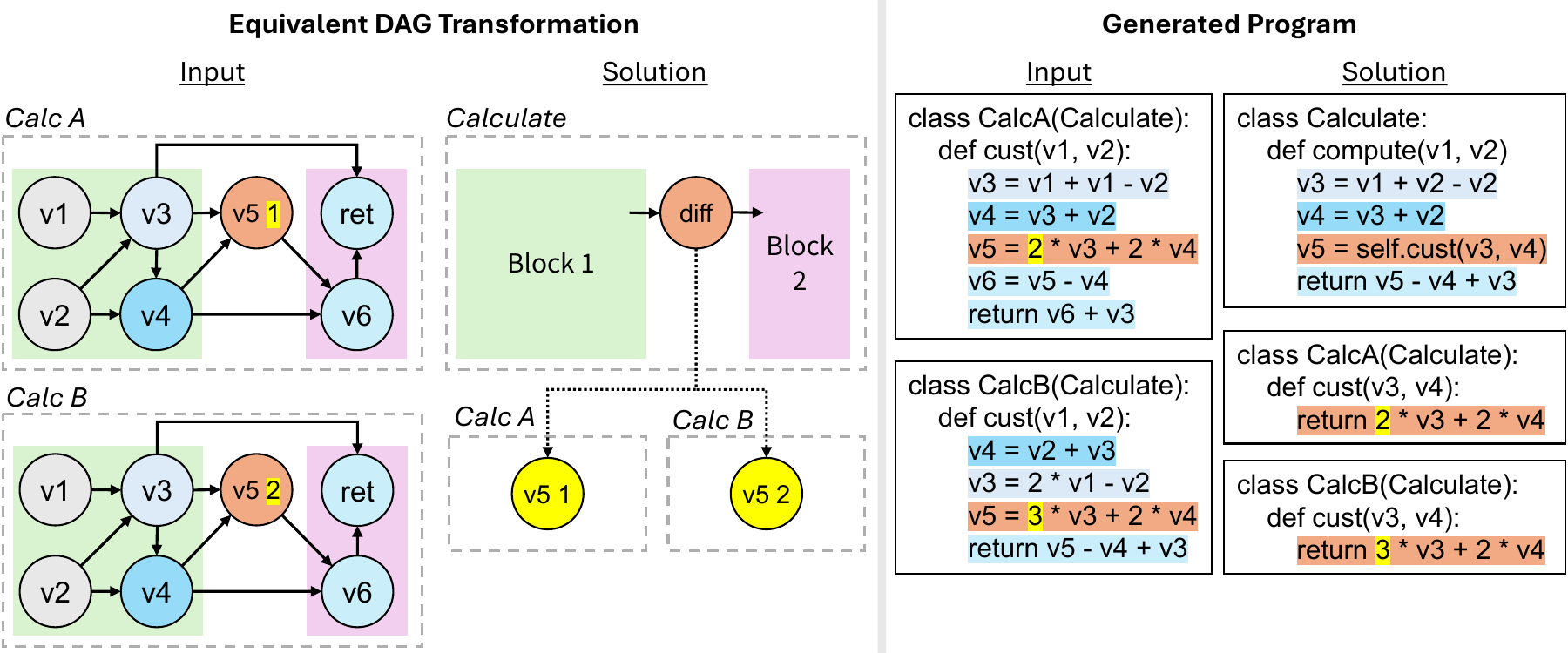}
    \caption{Our merge codebase task in \cref{subsection:understanding_task} can also be represented as a DAG transformation, mirroring the approach used in the original \model{} tasks (Figure~\ref{fig:cabra}, left). Each vertex $v$ represents a defined variable, with edge $v_i \rightarrow v_j$ indicating that variable $v_j$ uses $v_i$ in its computation (with randomly sampled operations). We then sample vertices $\mathcal{V}_\text{diff}$ to inject differences. To construct the ground-truth codebase $\mathcal{C}_\text{true}$, we move the subgraphs between each difference node into a parent class, and keep each difference node in the original classes.}
    \label{fig:merge_codebases_dag}
\end{figure}

\begin{figure}
    \centering
    \includegraphics[width=\linewidth]{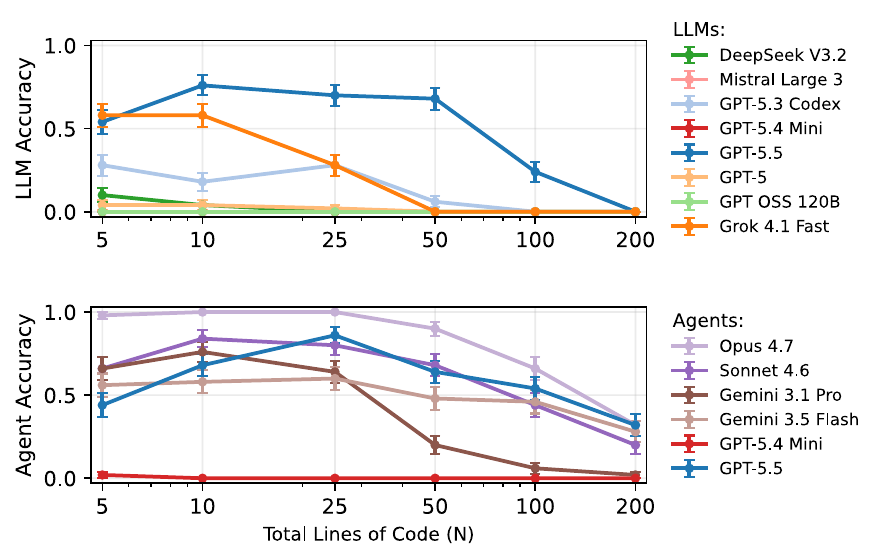}
    \caption{Comparison of LLM and Agent accuracy in the merge codebases task (\cref{section:understanding}). Both LLMs and agents struggle, with most LLMs failing to ever clear $0.5$ accuracy.
    }
    \label{fig:llm_and_agent_merge_code}
\end{figure}

\clearpage
\hypersetup{
    linkcolor=white,
    citecolor=white,
    urlcolor=white
}

\lstset{
  literate={<}{{<}}1
           {>}{{>}}1
}

\begin{prompt}[title={Prompt \thetcbcounter: Remove Dead Code Problem Statement}, label=prompt:remove_dead_code]
Some functions are unused when the solution.py script is run. Remove any function that is not reachable from main() through the call graph.

For example, if you were given the following code block:
<example>
import sys

def a(val: float) -> float:
    b_val = b(val)
    return b_val

def b(val: float) -> float:
    return val

def main(val: float):
    return b(val)

if __name__ == '__main__':
    main(float(sys.argv[1]))
</example>

You should remove the function `a()`, since it is not reached through `main()`, and leave `b()` as is:
<output>
import sys

def b(val: float) -> float:
    return val

def main(val: float):
    return b(val)

if __name__ == '__main__':
    main(float(sys.argv[1]))
</output>

You CANNOT add new functions, imports, or variables outside of the provided functions. You MUST preserve all function names that you do not remove (do not rename them). Do not change the "__main__" block or main() function.
\end{prompt}

\begin{prompt}[title={Prompt \thetcbcounter: Add Parameter Problem Statement}, label=prompt:add_parameter]
For any function in solution.py with the parameter mult, recursively propagate mult down the call graph by passing it to every function call used to compute the return expression. This will require adding mult to the parameter signatures of any called functions with a default value of `mult: float = default()`, and then recursively applying the same transformation to those functions. Continue propagating mult until reaching terminal functions whose return expressions directly compute a value `X` and do not call other functions in the file (apart from library functions). Only those terminal functions should apply the multiplier by returning mult * X; intermediate functions should NOT multiply their results by mult.

For example, if you were given the following code block:
<example>
import sys
from mylibrary import default

def a(val: float) -> float:
    b_val = b(val)
    return b_val

def b(val: float, mult: float = default()) -> float:
    c_val = c(val)
    return 2.0 * c_val

def c(val: float) -> float:
    return val + 0.5

def main(val: float):
    return a(val)

if __name__ == '__main__':
    main(float(sys.argv[1]))
</example>

You should modify the code such that a() is unchanged, and b() passes the `mult` parameter to c(). Then, you should modify the call signature of c() to accept `mult` with a default value of default(). Since c() is a terminal function, it will apply the multiplier to its original computation, returning `mult * (val + 0.5)`.
<output>
import sys
from mylibrary import default

def a(val: float) -> float:
    b_val = b(val)
    return b_val

def b(val: float, mult: float = default()) -> float:
    c_val = c(val, mult=mult)
    return 2.0 * c_val

def c(val: float, mult: float = default()) -> float:
    return mult * (val + 0.5)

def main(val: float):
    return a(val)

if __name__ == '__main__':
    main(float(sys.argv[1]))
</output>

You must follow this procedure exactly; adding function parameters or returning values that diverge from what we specify will cause your solution to be marked incorrect automatically.

Do NOT modify functions that are upstream of the function that already has the mult parameter. In the example above, a() is unchanged.

default() and mylibrary are undefined and will be defined later; do not try to define them. You CANNOT add new functions, imports, or variables outside of the provided functions. You MUST preserve all function names (do not rename them). You may define variables within functions. Do not change the "__main__" block or main() function. The code may change; do not assume that any default values will persist.
\end{prompt}

\begin{prompt}[title={Prompt \thetcbcounter: Add Return Value Problem Statement}, label=prompt:add_return]
In solution.py, external() previously returned a float, but mylibrary now defines it as returning tuple[float, str]. Update the code so the float preserves the original computation, while the str is propagated upward through the call chain. Any function that directly or indirectly uses the result of external() should return tuple[float, str]: the original computed value as the first element, and the propagated string as the second.

For example, if you were given the following code block:
<example>
import sys
from mylibrary import external

def a(val: float) -> float:
    b_val = b(val)
    c_val = c(val)
    return b_val + c_val

def b(val: float) -> float:
    return external(val)

def c(val: float) -> float:
    return val

def main(val: float):
    return a(val)

if __name__ == '__main__':
    main(float(sys.argv[1]))
</example>

You should modify the code such that b() now returns a tuple of type tuple[float, str] with the new values from external(). The string should propagate up the call graph recursively, modifying a() next, to preserve the original float computation. c() will remain unchanged.
<output>
import sys
from mylibrary import external

def a(val: float) -> tuple[float, str]:
    b_val, key = b(val)
    c_val = c(val)
    return b_val + c_val, key

def b(val: float) -> tuple[float, str]:
    return external(val)

def c(val: float) -> float:
    return val

def main(val: float):
    return a(val)

if __name__ == '__main__':
    main(float(sys.argv[1]))
</output>

You must follow this procedure exactly; adding function parameters or returning values that diverge from what we specify will cause your solution to be marked incorrect automatically.

If a function receives multiple propagated strings, it may return any one of them, since they are guaranteed to be identical. external() and mylibrary are undefined and will be defined later; do not try to define them. You CANNOT add new functions, imports, or variables outside of the provided functions. You MUST preserve all function names (do not rename them). You may define variables within functions. Do not change the "__main__" block or main() function.
\end{prompt}

\begin{prompt}[title={Prompt \thetcbcounter: Cache Function Problem Statement}, label=prompt:cache_function]
In solution.py, some functions call slow(), which is expensive and deterministic, so it should be run as few times as possible. Refactor the code so that each repeated call to slow() is performed exactly once at the lowest shared caller (i.e., lowest common ancestor) of the functions that need it. Store the result in a temporary variable called `temp`, then pass that value down as an additional argument to the functions that use it (e.g., a(val: float) -> a(val: float, temp: float)). 

For example, if you were given the following code block:
<example>
import sys
from mylibrary import slow

def a(val: float) -> float:
    b_val = b(val)
    return b_val

def b(val: float) -> float:
    c_val = c(val)
    d_val = d(val)
    return c_val + d_val

def c(val: float) -> float:
    return -4 * slow()

def d(val: float) -> float:
    e_val = e(val)
    return 2 * slow() + e_val

def e(val: float) -> float:
    return val

def main(val: float):
    return a(val)

if __name__ == '__main__':
    main(float(sys.argv[1]))
</example>

You should modify the code such that the calls to slow() in c() and d() are moved to b(), as b() is the lowest common ancestor of c() and d(). Then, b() stores the intermediate value of slow() and passes it down to the other functions:
<output>
import sys
from mylibrary import slow

def a(val: float) -> float:
    b_val = b(val)
    return b_val

def b(val: float) -> float:
    temp: float = slow()
    c_val = c(val, temp)
    d_val = d(val, temp)
    return c_val + d_val

def c(val: float, temp: float) -> float:
    return -4 * temp

def d(val: float, temp: float) -> float:
    e_val = e(val)
    return 2 * temp + e_val

def e(val: float) -> float:
    return val

def main(val: float):
    return a(val)

if __name__ == '__main__':
    main(float(sys.argv[1]))
</output>
\end{prompt}

\begin{prompt}[title={Prompt \thetcbcounter: Don't Repeat Yourself Problem Statement}, label=prompt:dont_repeat]
In solution.py, multiple functions independently call get_multiplier(val), which returns a multiplier for a given value. To reduce this duplication, refactor the code so the multiplier is defined once at the earliest shared downstream function reached by following calls from each of those functions, then return that multiplier back up to the functions that need it. Do not move the multiplier computation into a caller/ancestor of the duplicated functions.

For example, if you were given the following code block:
<example>
import sys
from mylibrary import get_multiplier

def a(val: float) -> float:
    b_val = b(val)
    c_val = c(val)
    return b_val + c_val

def b(val: float) -> float:
    d_val = d(val)
    e_val = e(val)
    return get_multiplier(val) * (d_val + e_val)

def c(val: float) -> float:
    d_val = d(val)
    return get_multiplier(val) * d_val

def d(val: float) -> float:
    e_val = e(val)
    return e_val

def e(val: float) -> float:
    return val

def main(val: float):
    return a(val)

if __name__ == '__main__':
    main(float(sys.argv[1]))
</example>

First, you should identify that b() and c() both call get_multiplier(val). Then, trace their calls downward toward their callees. The first shared downstream function is d(), so move get_multiplier(val) into d(), return it along with d()'s original result, and reuse that returned multiplier in b() and c().
<output>
import sys
from mylibrary import get_multiplier

def a(val: float) -> float:
    b_val = b(val)
    c_val = c(val)
    return b_val + c_val

def b(val: float) -> float:
    d_val, mult = d(val)
    e_val = e(val)
    return mult * (d_val + e_val)

def c(val: float) -> float:
    d_val, mult = d(val)
    return mult * d_val

def d(val: float) -> tuple[float, float]:
    e_val = e(val)
    return e_val, get_multiplier(val)

def e(val: float) -> float:
    return val

def main(val: float):
    return a(val)

if __name__ == '__main__':
    main(float(sys.argv[1]))
</output>
\end{prompt}

\begin{prompt}[title={Prompt \thetcbcounter: Merge Codebases Problem Statement}, label=prompt:merge_code]
CalculateA and CalculateB are subclasses of the shared Calculate class. Their compute() methods contain redundant logic. Refactor the code using the template method pattern: move as much identical compute() logic as possible into Calculate.compute(), extract subclass-specific behavior into helper methods with the name custom_compute_{idx}, and keep only the differences in CalculateA and CalculateB.

For example, if you were given the following code block:
<example>
class Calculate:
    pass

class CalculateA(Calculate):
    def compute(self, a: int) -> int:
        b = a * 2
        c = b + 3
        return c * 10

class CalculateB(Calculate):
    def compute(self, c: int) -> int:
        a = 2 * c
        b = a - 1
        return b * 10
</example>

You should modify the code such that Calculate.compute() contains the shared control flow and shared operations. The child classes should only implement the custom part that differs between them:
<output>
class Calculate:
    def compute(self, val_0: int) -> int:
        val_1 = val_0 * 2
        val_2 = self.custom_compute_1(val_1)
        return val_2 * 10

class CalculateA(Calculate):
    def custom_compute_1(self, val: int) -> int:
        return val + 3

class CalculateB(Calculate):
    def custom_compute_1(self, val: int) -> int:
        return val - 1
</output>

You may need to repeat this process multiple times with multiple custom_comput_{idx}() functions if there are multiple identical regions and multiple smaller differences inside compute(). The final Calculate.compute() method should contain all shared logic and call child class-specific helper methods only where CalculateA and CalculateB differ.

You must preserve the behavior of CalculateA.compute() and CalculateB.compute(). You MUST keep CalculateA and CalculateB as subclasses of Calculate. You CANNOT add new imports or unrelated classes. You may add helper methods to Calculate, CalculateA, and CalculateB as needed. Do not rename Calculate, CalculateA, CalculateB, or any public methods used outside these classes. Keep as few lines of code as possible in the child class functions custom_compute_{idx}; only subclass-specific differences should remain there.
\end{prompt}
\begin{prompt}[title={Prompt \thetcbcounter: Add Parameter Problem Statement --- String Variation}, label=prompt:add_parameter_string]
For any function in solution.py with the parameter prefix, recursively propagate prefix down the call graph by passing it to every function call used to compute the return expression. This will require adding prefix to the parameter signatures of any called functions with a default value of `prefix: str = default()`, and then recursively applying the same transformation to those functions. Continue propagating prefix until reaching terminal functions whose return expressions directly compute a value `X` and do not call other functions in the file (apart from library functions). Only those terminal functions should add the prefix by returning prefix + X; intermediate functions should NOT append the prefix to their results.

For example, if you were given the following code block:
<example>
import sys
import strops
from mylibrary import default

def a(val: str) -> str:
    b_val = b(val)
    return b_val

def b(val: str, prefix: str = default()) -> str:
    c_val = c(val)
    return strops.reverse(c_val)

def c(val: str) -> str:
    return strops.interleave(val, "abc")

def main(val: str):
    return a(val)

if __name__ == '__main__':
    main(str(sys.argv[1]))
</example>

You should modify the code such that a() is unchanged, and b() passes the `prefix` parameter to c(). Then, you should modify the call signature of c() to accept `prefix` with a default value of default(). Since c() is a terminal function, it will apply the prefix to its original computation, returning `prefix + strops.interleave(val, "abc")`.
<output>
import sys
import strops
from mylibrary import default

def a(val: str) -> str:
    b_val = b(val)
    return b_val

def b(val: str, prefix: str = default()) -> str:
    c_val = c(val, prefix=prefix)
    return strops.reverse(c_val)

def c(val: str, prefix: str = default()) -> str:
    return prefix + strops.interleave(val, "abc")

def main(val: str):
    return a(val)

if __name__ == '__main__':
    main(str(sys.argv[1]))
</output>

You must follow this procedure exactly; adding function parameters or returning values that diverge from what we specify will cause your solution to be marked incorrect automatically.

Do NOT modify functions that are upstream of the function that already has the prefix parameter. In the example above, a() is unchanged.

strops, default(), and mylibrary are undefined and will be defined later; do not try to define them. You CANNOT add new functions, imports, or variables outside of the provided functions. You MUST preserve all function names (do not rename them). You may define variables within functions. Do not change the "__main__" block or main() function. The code may change; do not assume that any default values will persist.
\end{prompt}

\begin{prompt}[title={Prompt \thetcbcounter: Add Parameter Problem Statement --- Array Variation}, label=prompt:add_parameter_array]
For any function in solution.py with the parameter mult, recursively propagate mult down the call graph by passing it to every function call used to compute the return expression. This will require adding mult to the parameter signatures of any called functions with a default value of `mult: float = default()`, and then recursively applying the same transformation to those functions. Continue propagating mult until reaching terminal functions whose return expressions directly compute a value `X` and do not call other functions in the file (apart from library functions). Only those terminal functions should apply the multiplier by returning mult * X; intermediate functions should NOT multiply their results by mult.

For example, if you were given the following code block:
<example>
import sys
import numpy as np
from mylibrary import default

def a(val: np.ndarray) -> np.ndarray:
    b_val = b(val)
    return b_val

def b(val: np.ndarray, mult: float = default()) -> np.ndarray:
    c_val = c(val)
    return 2.0 * c_val

def c(val: np.ndarray) -> np.ndarray:
    return val + 0.5

def main(val: np.ndarray):
    return a(val)

if __name__ == '__main__':
    main(np.array(eval(sys.argv[1])))
</example>

You should modify the code such that a() is unchanged, and b() passes the `mult` parameter to c(). Then, you should modify the call signature of c() to accept `mult` with a default value of default(). Since c() is a terminal function, it will apply the multiplier to its original computation, returning `mult * (val + 0.5)`.
<output>
import sys
import numpy as np
from mylibrary import default

def a(val: np.ndarray) -> np.ndarray:
    b_val = b(val)
    return b_val

def b(val: np.ndarray, mult: float = default()) -> np.ndarray:
    c_val = c(val, mult=mult)
    return 2.0 * c_val

def c(val: np.ndarray, mult: float = default()) -> np.ndarray:
    return mult * (val + 0.5)

def main(val: np.ndarray):
    return a(val)

if __name__ == '__main__':
    main(np.array(eval(sys.argv[1])))
</output>

You must follow this procedure exactly; adding function parameters or returning values that diverge from what we specify will cause your solution to be marked incorrect automatically.

Do NOT modify functions that are upstream of the function that already has the mult parameter. In the example above, a() is unchanged.

default() and mylibrary are undefined and will be defined later; do not try to define them. You CANNOT add new functions, imports, or variables outside of the provided functions. You MUST preserve all function names (do not rename them). You may define variables within functions. Do not change the "__main__" block or main() function. The code may change; do not assume that any default values will persist.
\end{prompt}
\begin{prompt}[title={Prompt \thetcbcounter: Add Parameter Task --- Scaling Runtime (See Code Block~\ref{code:collatz})}, label=prompt:runtime]
For any function in solution.py with the parameter mult, recursively propagate mult down the call graph by passing it to every function call used to compute the return expression. This will require adding mult to the parameter signatures of any called functions with a default value of `mult: float = default()`, and then recursively applying the same transformation to those functions. Continue propagating mult until reaching terminal functions whose return expressions directly compute a value `X` and do not call other functions in the file (apart from library functions). Only those terminal functions should apply the multiplier by returning mult * X; intermediate functions should NOT multiply their results by mult. Some code may be guarded by conditional expressions; only the if/else branch that actually executes influences the output. Apply this transformation only inside branches that actually run; if another branch contains the same calls or computation, preserve it unchanged.

For example, if you were given the following code block:
<example>
import sys
from mylibrary import default

def check():
    return True

def a(val: float) -> float:
    b_val = b(val)
    return b_val

def b(val: float, mult: float = default()) -> float:
    if check():
        c_val = c(val)
        return 2.0 * c_val
    else:
        c_val = c(val)
        return 2.0 * c_val

def c(val: float) -> float:
    return val + 0.5

def main(val: float):
    return a(val)

if __name__ == '__main__':
    main(float(sys.argv[1]))
</example>

You should modify the code such that a() is unchanged, and b() passes the `mult` parameter to c() only in the branch that actually executes. Then, you should modify the call signature of c() to accept `mult` with a default value of default(). Since c() is a terminal function on the running branch, it will apply the multiplier to its original computation, returning `mult * (val + 0.5)`. Do not pass `mult` in the branch that does not run, even though it contains the same computation.
<output>
import sys
from mylibrary import default

def check():
    return True

def a(val: float) -> float:
    b_val = b(val)
    return b_val

def b(val: float, mult: float = default()) -> float:
    if check():
        c_val = c(val, mult=mult)
        return 2.0 * c_val
    else:
        c_val = c(val)
        return 2.0 * c_val

def c(val: float, mult: float = default()) -> float:
    return mult * (val + 0.5)

def main(val: float):
    return a(val)

if __name__ == '__main__':
    main(float(sys.argv[1]))
</output>

...
\end{prompt}

\begin{prompt}[title={Prompt \thetcbcounter: Add Parameter Task --- Scaling Instruction Following}, label=prompt:instructions]
...

In addition to the propagation above, each generated parameter carries a set of CONSTRAINTS that are applied in every eligible function. The constraints for all generated parameters are listed together in the "requirements" section below and may be shuffled rather than grouped by parameter. Apply these constraints inside every function that must take the named parameter as part of the propagation. For each such function, apply every requirement for the parameter named in that requirement. Do not apply a requirement inside main(), inside functions that do not take the named parameter.

When several requirements match the same function, apply them in this fixed order within the function body: 1. comments 2. assert statements 3. print statements 4. temporary-variable assignments

Insert these constraint lines after the function has received the named parameter and before the named parameter is passed to another function or used in a terminal return computation. Comment requirements should be written as real # comments on their own lines, exactly as specified, including the named generated value. Assert, print, and temporary-variable assignment requirements should be written as separate Python statements on their own lines.

Temporary-variable assignment requirements compute a value from the named parameter and store it in the exact temporary variable named by the requirement. They do not update the parameter itself. For example, if a requirement says to set mult1_temp to mult1 + 0.2, write mult1_temp = mult1 + 0.2; continue passing and using mult1, not mult1_temp. If there are multiple matching requirements of the same kind, write each one as a separate line.

Here are the 10 requirements for the mult parameters (mult1, mult2, mult3, mult4, mult5, mult6, mult7, mult8, mult9, and mult10). These requirements apply to every eligible function.

1. For mult8: In every function, assert() that mult8 is not equal to 30.
2. For mult7: In every function, assert() that mult7 differs from 30.
3. For mult6: In every function, print mult6 prefixed with the string 'Log: ' in an fstring.
4. For mult4: In every function, assert() that mult4 is not equal to -30.
5. For mult8: In every function, add the comment '# TODO: mult8' in the function.
6. For mult7: In every function, compute mult7/0.4 into snapshot_mult7 on a separate line.
7. For mult5: In every function, make sure mult5 doesn't equal -30 via an assert() statement.
8. For mult2: In every function, add the exact comment '# verify: mult2' to the function.
9. For mult2: In every function, compute mult2-0.7 into mult2_local on a separate line.
10. For mult4: In every function, assert() that mult4 is not equal to -10.
\end{prompt}

\begin{prompt}[title={Prompt \thetcbcounter: Add Parameter Task --- Comment Copying}, label=prompt:instructions_edit_only]
...

For every function you edit, add this comment on a new line verbatim: "# buoyancy victory majesty fibre audible lancer lovemaking fathom readability proletarian bathroom jogging copernicus fight nutrition thatcher subset search means nowadays cochin prologue entrails reissue cyberspace"
\end{prompt}

\begin{prompt}[title={Prompt \thetcbcounter: Variable Replacement Task}, label=prompt:replace_vars]
solution.py has a single function called main(val: float) that performs a math operation using intermediate variables. Rename the variables as follows:
- Change the name of variable 'gg' to 'bt'
- Change the name of variable 'cn' to 'gi'
- Change the name of variable 'fv' to 'g'
- Change the name of variable 'd' to 'ac'
- Change the name of variable 'dh' to 'fi'

You MUST apply every requested rename consistently to assignments and uses. You MUST NOT rename any other variables or change the code in any other way.
\end{prompt}
\begin{prompt}[title={Prompt \thetcbcounter: LLM Generation --- Full Response}, label=prompt:llm_full]
<instructions>
You are a helpful assistant that solves programming tasks.

You will be given an initial code repository and a problem statement. Your goal is to modify the repository to solve the problem.

Your response must contain exactly ONE <solution> code block with your solution to the problem.
Include a <thought> section before your <solution> where you explain your reasoning process.
Format your response as shown in <format_example>.

<format_example>
<thought>
Your reasoning and analysis here. Explain why you came up with this solution
</thought>

<solution>
Your solution to the problem. Return the entire modified repository in full in the format below:

# === solution.py (any file name) ===
...[insert code]...

This would correspond to writing all of `solution.py`.

</solution>
</format_example>

Failure to follow these rules will cause your response to be rejected.
</instructions>

Here is the initial code repository you must modify:
<code repository>
{code_repository}
</code repository>

Here is the problem statement you must satisfy:
<problem statement>
{problem_statement}
</problem statement>
\end{prompt}

\begin{prompt}[title={Prompt \thetcbcounter: LLM Generation --- Diff Patch}, label=prompt:llm_diff]
...

<format_example>

<thought>
Your reasoning and analysis here. Explain why you came up with this solution
</thought>

<solution>
Your solution to the problem. For each file that needs to be changed, write out the changes similar to a unified diff like `diff -U0` would produce. Start each hunk of changes with a `@@ ... @@` line. Don't include line numbers like `diff -U0` does.

Return your modifications in the diff format shown below:

```diff
--- app.py
+++ app.py
@@ ... @@
-class MathWeb:
+import sympy
+
+class MathWeb:
@@ ... @@
-def is_prime(x):
-    if x < 2:
-        return False
-    for i in range(2, int(math.sqrt(x)) + 1):
-        if x % i == 0:
-            return False
-    return True
@@ ... @@
-@app.route('/prime/<int:n>')
-def nth_prime(n):
-    count = 0
-    num = 1
-    while count < n:
-        num += 1
-        if is_prime(num):
-            count += 1
-    return str(num)
+@app.route('/prime/<int:n>')
+def nth_prime(n):
+    count = 0
+    num = 1
+    while count < n:
+        num += 1
+        if sympy.isprime(num):
+            count += 1
+    return str(num)
```
    
This would correspond to replacing the function call to `is_prime` with a call to sympy in a file called app.py
</solution>

Overall, remember to format your solution as:
<thought>
...
</thought>
<solution>
...
</solution>

</format_example>

Failure to follow these rules will cause your response to be rejected.
</instructions>

...
\end{prompt}
\begin{prompt}[title={Prompt \thetcbcounter: Classifying Tool Calls on \model{}}, label=prompt:classify_cabra]
<task>
You are an expert at classifying the reason that a coding agent generated a script or tool call for completing a programming task.

You will be given information about an agent's tool call, including the content of the tool call and a summary of what the tool call does. Your job is to classify the tool call into one of the following categories:
1. Read: Call view or a similar tool to list all of the files/folders in the codebase, review the entire codebase, look at an entire file, or view a certain section of the file (i.e., span of lines). The point is we are mostly reading the code exactly without doing analysis on it
2. Search: Use a search tool such as `grep`, `find`, `glob`, or `rg` to look for a certain parameter, function, or snippet in the code. The point is we are searching for something specifically that exists exactly in the code.
3. Understand: Use a script (e.g., Abstract Syntax Tree) to comprehend the current codebase, including dependencies, call graph analysis, and runtime analysis (such as checking the parity returned by the `check()` output function). The point is we are doing analysis on top of the code beyond just reading or searching. This also includes creating new files/scripts to do this analysis.
4. Edit: Make edits to the code in solution.py, including any initial changes or any revisions (e.g., based on failed tests). This also includes the generation of meta-scripts that can perform code edits after the script is run
5. Test: Test the current code, including syntax validation, test case generation, mock executions, or smoke tests. This also includes making changes or creating files that are not solution.py (e.g., create test.py file)
6. Summary: The model calls no tool and instead just generates a summary of the changes made, as the task is complete. The tool_name will be "task_complete"
7. Plan: Create a plan (usually with SQL) or TODOs with steps on what to do
8. Other: Any other change that is not described above (e.g., web search, removing files, package installations)
</task>

If the model creates a new file, you need to determine what is the intent of this script and classify it accordingly (typically: understand, edit, or test).

Here is the input tool call and message information that you must classify:
<tool call information>
{tool_call}
</tool call information>

<format>
Return only a single word with your classification, which must be one of the following labels:
- read
- search
- understand
- edit
- test
- summary
- plan
- other

Do not generate ANYTHING else
</format>
\end{prompt}

\begin{prompt}[title={Prompt \thetcbcounter: Classifying Tool Calls on SWE-Bench Verified}, label=prompt:classify_swebench]
<task>
You are an expert at classifying the reason that a coding agent generated a script or tool call for completing a programming task.

You will be given information about an agent's tool call, including the content of the tool call and a summary of what the tool call does. Your job is to classify the tool call into one of the following categories:
1. Read: Call view or a similar tool to list all of the files/folders in the codebase, review the entire codebase, look at an entire file, or view a certain section of the file (i.e., span of lines). The point is we are mostly reading the code exactly without doing analysis on it
2. Search: Use a search tool such as `grep`, `find`, `glob`, or `rg` to look for a certain parameter, function, or snippet in the code. The point is we are searching for something specifically that exists exactly in the code.
3. Understand: Use a script (e.g., Abstract Syntax Tree) to comprehend the current codebase, including dependencies, call graph analysis, and runtime analysis (e.g., checking the output of a function). The point is we are doing analysis on top of the code beyond just reading or searching. This also includes creating new files/scripts to do this analysis.
4. Edit: Make edits to the code, including any initial changes or any revisions (e.g., based on failed tests). This also includes the generation of meta-scripts that can perform code edits after the script is run.
5. Test: Test the current code or reproduce the current error, including syntax validation, test case generation/running, mock executions, or smoke tests. This also includes making changes or creating files used for testing.
6. Summary: The model calls no tool and instead just generates a summary of the changes made, as the task is complete. The tool_name will similar to "task_complete"
7. Plan: Create a plan with explicit TODOs on what to do
8. Other: Any other change that is not described above (e.g., web search, removing files, package installations)
</task>

If the model creates a new file, you need to determine what is the intent of this script and classify it accordingly (typically: understand, edit, or test).

Here is the input tool call and message information that you must classify:
<tool call information>
{tool_call}
</tool call information>

<format>
Return only a single word with your classification, which must be one of the following labels:
- read
- search
- understand
- edit
- test
- summary
- plan
- other

Do not generate ANYTHING else
</format>
\end{prompt}

\begin{codeblock}[title={Code Block \thetcbcounter: Example Math Program},
  label={code:math}]
from mylibrary import default
import math
import operator
import sys

def n(val: float) -> float:
    return abs(val)

def d(val: float) -> float:
    g_val = g(val)
    r_val = r(val)
    return operator.add(g_val, r_val)

def b(val: float) -> float:
    return abs(val)

def q(val: float) -> float:
    b_val = b(val)
    f_val = f(val)
    a_val = a(val)
    j_val = j(val)
    return operator.sub(min(math.hypot(b_val, f_val), a_val), j_val)

def s(val: float, mult: float=default()) -> float:
    q_val = q(val)
    return round(q_val)

def g(val: float) -> float:
    l_val = l(val)
    i_val = i(val)
    return max(l_val, i_val)

def i(val: float) -> float:
    return math.degrees(val)

def a(val: float) -> float:
    return operator.neg(val)

def j(val: float) -> float:
    return math.cos(val)

def f(val: float) -> float:
    return abs(val)

def l(val: float) -> float:
    return math.tanh(val)

def r(val: float) -> float:
    n_val = n(val)
    i_val = i(val)
    s_val = s(val)
    return math.hypot(math.hypot(n_val, i_val), s_val)

def main(val: float):
    return d(val)

if __name__ == '__main__':
    main(float(sys.argv[1]))
\end{codeblock}

\begin{codeblock}[title={Code Block \thetcbcounter: Example String Program},
  label={code:string}]
from mylibrary import default
import strops
import sys

def n(val: str) -> str:
    return strops.rot13(val)

def d(val: str) -> str:
    g_val = g(val)
    r_val = r(val)
    return strops.interleave(g_val, r_val)

def b(val: str) -> str:
    return strops.rot13(val)

def q(val: str) -> str:
    b_val = b(val)
    f_val = f(val)
    a_val = a(val)
    j_val = j(val)
    return strops.swap_halves(strops.zip_chars(strops.xor_chars(b_val, f_val), a_val), j_val)

def s(val: str, prefix: str=default()) -> str:
    q_val = q(val)
    return strops.mirror(q_val)

def g(val: str) -> str:
    l_val = l(val)
    i_val = i(val)
    return strops.alternate_chars(l_val, i_val)

def i(val: str) -> str:
    return strops.char_rotate(val)

def a(val: str) -> str:
    return strops.char_rotate_by_3(val)

def j(val: str) -> str:
    return strops.reverse(val)

def f(val: str) -> str:
    return strops.rot13(val)

def l(val: str) -> str:
    return strops.swap_case(val)

def r(val: str) -> str:
    n_val = n(val)
    i_val = i(val)
    s_val = s(val)
    return strops.xor_chars(strops.xor_chars(n_val, i_val), s_val)

def main(val: str):
    return d(val)

if __name__ == '__main__':
    main(str(sys.argv[1]))
\end{codeblock}

\begin{codeblock}[title={Code Block \thetcbcounter: Example Array Program},
  label={code:array}]
from mylibrary import default
import numpy as np
import sys

def n(val: np.ndarray) -> np.ndarray:
    return np.abs(val)

def d(val: np.ndarray) -> np.ndarray:
    g_val = g(val)
    r_val = r(val)
    return np.add(g_val, r_val)

def b(val: np.ndarray) -> np.ndarray:
    return np.abs(val)

def q(val: np.ndarray) -> np.ndarray:
    b_val = b(val)
    f_val = f(val)
    a_val = a(val)
    j_val = j(val)
    return np.subtract(np.multiply(np.minimum(b_val, f_val), a_val), j_val)

def s(val: np.ndarray, mult: float=default()) -> np.ndarray:
    q_val = q(val)
    return np.sign(q_val)

def g(val: np.ndarray) -> np.ndarray:
    l_val = l(val)
    i_val = i(val)
    return np.maximum(l_val, i_val)

def i(val: np.ndarray) -> np.ndarray:
    return np.cumprod(val)

def a(val: np.ndarray) -> np.ndarray:
    return np.flip(val)

def j(val: np.ndarray) -> np.ndarray:
    return np.sin(val)

def f(val: np.ndarray) -> np.ndarray:
    return np.abs(val)

def l(val: np.ndarray) -> np.ndarray:
    return np.cos(val)

def r(val: np.ndarray) -> np.ndarray:
    n_val = n(val)
    i_val = i(val)
    s_val = s(val)
    return np.minimum(np.minimum(n_val, i_val), s_val)

def main(val: np.ndarray):
    return d(val)

if __name__ == '__main__':
    main(np.array(eval(sys.argv[1])))
\end{codeblock}

\begin{codeblock}[title={Code Block \thetcbcounter: Collatz-Inspired Runtime Function},
  label={code:collatz}]
def check(x: int, a: int, b: int, target: int, num_steps: int) -> bool:
    for step in range(num_steps):
        if x % 2 == 0:
            x = x // 2
        else:
            x = a * x + b

    return x % 2 == target
\end{codeblock}

\begin{codeblock}[title={Code Block \thetcbcounter: Example Calculator Classes},
  label={code:calculator}]
from abc import ABC, abstractmethod


class Calculate(ABC):
    """Abstract base: turn the input parameters into a single result."""

    @abstractmethod
    def compute(self, *args):
        ...


class CalculateA(Calculate):

    def compute(self, al, ac, ae, aj, l, ak, ag, m, ad, ab):

        af = 4 * ac - 2 * (aj * l)
        s = -5 * (ag * af) - aj - aj
        v = 2 * s - ab

        ah = -4 * ac + v

        y = -4 * (aj * s) - 2 * ac
        q = y - 2 * af
        x = 5 * ah + 4 * ac

        return x + q

class CalculateB(Calculate):

    def compute(self, l, aa, ab, t, o, ah, w, y, v, r):

        aj = 4 * aa - 2 * (o * t)
        s = -2 * t - 5 * (w * aj)
        al = -r + 2 * s

        ai = -4 * (aa * v) + al

        af = 4 * aa + 5 * ai
        m = -4 * (t * s) - 2 * aa
        z = -2 * aj + m

        return af + z


if __name__ == "__main__":
    calc_a = CalculateA()
    calc_b = CalculateB()

    args = (1, 2, 3, 4, 5, 6, 7, 8, 9, 10)

    result_a = calc_a.compute(*args)
    result_b = calc_b.compute(*args)

    print(f"Result A: {result_a}")
    print(f"Result B: {result_b}")
\end{codeblock}

\begin{codeblock}[title={Code Block \thetcbcounter: Rename Variable Program},
  label={code:rename_variables}]
def main(val: float):
    ak = val
    n = ak + 4
    gr = n + ak
    ds = gr + ak
    bk = ds - ak
    f = bk - n
    bn = f + 4
    ae = bn + 8
    bb = ae + 7
    dz = bb - bn
    bj = dz + ae
    cd = bj + 2
    cm = cd + f
    gk = cm - ds
    al = gk - gr
    ft = al + gk
    dg = ft + gk
    at = dg + ak
    p = at + 4
    gq = p + ft
    c = gq + 1
    ee = c - cm
    ff = ee - 8
    aq = ff - at
    dr = aq + dz
    fx = dr + 9
    fw = fx + 1
    gx = fw + 9
    dy = gx + dg
    fk = dy + 2
    d = fk - 4
    cc = d + dy
    ed = cc + 1
    bw = ed - c
    cz = bw + c
    fc = cz - bb
    cb = fc + bk
    fu = cb + bn
    fh = fu - 9
    j = fh + ff
    m = j - 6
    am = m - cz
    aw = am + 9
    au = aw - 9
    ch = au + 2
    cx = ch + bn
    bz = cx + fh
    return bz
\end{codeblock}

\end{document}